\documentclass[twocolumn,prd,superscriptaddress,showpacs,floatfix,%
preprintnumbers,nofootinbib]{revtex4}

\usepackage{epsfig}
\usepackage{bm}

\begin{document}

\title{Adiabaticity and spectral splits in
collective neutrino transformations}

\author{Georg G.~Raffelt}
\affiliation{Max-Planck-Institut f\"ur Physik
(Werner-Heisenberg-Institut), F\"ohringer Ring 6, 80805 M\"unchen,
Germany}

\author{Alexei Yu.~Smirnov}
\affiliation{Abdus Salam International Centre for Theoretical Physics,
Strada Costiera 11, 34014 Trieste, Italy}
\affiliation{Institute for Nuclear Research, Russian Academy of
Sciences,
60th October Anniversary Prospect 7A, 117 312 Moscow, Russia}

\date{28 September 2007}

\preprint{MPP-2007-120}

\begin{abstract}
Neutrinos streaming off a supernova core transform collectively by
neutrino-neutrino interactions, leading to ``spectral splits'' where
an energy $E_{\rm split}$ divides the transformed spectrum sharply
into parts of almost pure but different flavors.  We present a
detailed description of the spectral split phenomenon which is
conceptually and quantitatively understood in an adiabatic treatment
of neutrino-neutrino effects. Central to this theory is a
self-consistency condition in the form of two sum rules (integrals
over the neutrino spectra that must equal certain conserved
quantities).  We provide explicit analytic and numerical solutions for
various neutrino spectra.  We introduce the concept of the adiabatic
reference frame and elaborate on the relative adiabatic evolution.
Violating adiabaticity leads to the spectral split being ``washed
out.''  The sharpness of the split appears to be represented by a
surprisingly universal function.
\end{abstract}


\maketitle

\section{Introduction}                        \label{sec:introduction}

In a dense neutrino gas, neutrino-neutrino refraction causes
nonlinear flavor oscillation phenomena~\cite{Pantaleone:1992eq,
Samuel:1993uw, Samuel:1996ri, Qian:1995ua, Fuller:2005ae,
Pastor:2001iu, Pastor:2002we, Duan:2005cp, Duan:2006an, Duan:2006jv,
Hannestad:2006nj, Duan:2007mv, Raffelt:2007yz, Raffelt:2007cb,
EstebanPretel:2007ec, Duan:2007fw, Duan:2007bt, Fogli:2007bk}. In the
region between the neutrino sphere and a radius of several hundred
kilometers in a core-collapse supernova (SN), the flavor content of
neutrino fluxes is dramatically modified~\cite{Pastor:2002we,
Duan:2005cp, Duan:2006an, Duan:2006jv, Hannestad:2006nj, Duan:2007mv,
Raffelt:2007yz, Raffelt:2007cb, EstebanPretel:2007ec, Duan:2007fw,
Duan:2007bt, Fogli:2007bk}. Ordinary MSW resonances typically occur
at much larger distances and process the spectra further in
well-understood ways~\cite{Dighe:1999bi, Dighe:2004xy}.

\begin{figure}[b]
\includegraphics[width=0.85\columnwidth]{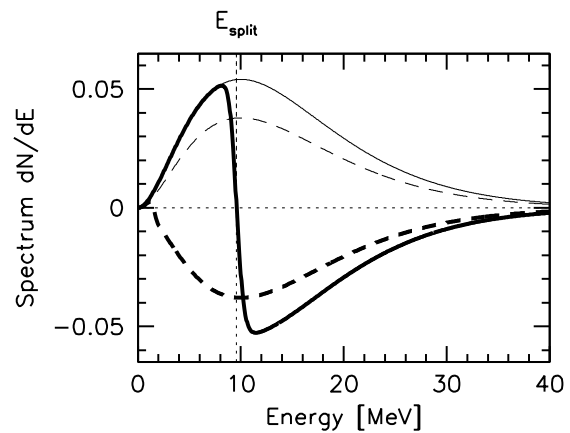}
\caption{Neutrino spectra at the neutrino sphere (thin lines) and
beyond the dense-neutrino region (thick lines) for the schematic SN
model described in the text. Solid: neutrinos. Dashed:
antineutrinos. Positive spectrum: electron (anti)neutrinos.
Negative spectrum: $x$ (anti)neutrinos.\label{fig:snspectrum}}
\end{figure}

One particularly intriguing feature of the emerging fluxes is a
``spectral split.'' What this means is best described with the help of
an example. In Fig.~\ref{fig:snspectrum} we show thermal $\nu_e$ and
$\bar\nu_e$ flux spectra with an average energy of 15~MeV produced in
the neutrino sphere of a SN core (thin lines). This example is
schematic in that we assume equal average $\nu_e$ and $\bar\nu_e$
energies but an overall $\bar\nu_e$ flux that is only 70\% of the
$\nu_e$ flux. Moreover, the smaller fluxes of the other species
$\nu_\mu$, $\nu_\tau$, $\bar\nu_\mu$ and $\bar\nu_\tau$ are completely
ignored. We are considering two-flavor oscillations between $\nu_e$
and another flavor $\nu_x$, driven by the atmospheric mass squared
difference $\Delta m^2=2$--$3\times10^{-3}~{\rm eV}^2$ and the small
1-3 mixing angle. Therefore in Fig.~\ref{fig:snspectrum} we show the
$z$--components of the usual two-flavor polarization vectors, where
``up'' denotes the electron flavor and ``down'' the $x$--flavor. In
other words, the spectrum represents electron (anti)neutrinos where it
is positive and $x$ (anti)neutrinos where it is negative.

After a few 100~km, the neutrino-neutrino effects have completely died
out. The emerging spectra are shown as thick lines. The antineutrinos
(dashed) have completely flipped to the $x$--flavor.  The same is true
for the neutrinos (solid) down to a critical energy $E_{\rm split}$
where the spectrum splits. All $\nu_e$ below this energy emerge in
their original flavor. Spectral splits of this sort were first
observed in the numerical simulations of Duan et
al.~\cite{Duan:2006an} where a first interpretation was given. Later
these authors have introduced the term ``stepwise spectral swapping''
to describe this phenomenon~\cite{Duan:2007bt}.

In a previous short paper~\cite{Raffelt:2007cb} we have shown that
$E_{\rm split}$ is fixed by what we call lepton-number conservation.
For a sufficiently small in-medium mixing angle the collective flavor
transformations have the property of conserving flavor number in the
sense that only pairs $\nu_e\bar\nu_e$ are transformed to pairs
$\nu_x\bar\nu_x$, whereas the excess $\nu_e$ flux is
conserved~\cite{Hannestad:2006nj}. While this property does not
explain the existence of a spectral split, it explains its location
that for the example of Fig.~\ref{fig:snspectrum} is at $E_{\rm
split}=9.57~{\rm MeV}$.

\begin{figure}
\includegraphics[width=0.75\columnwidth]{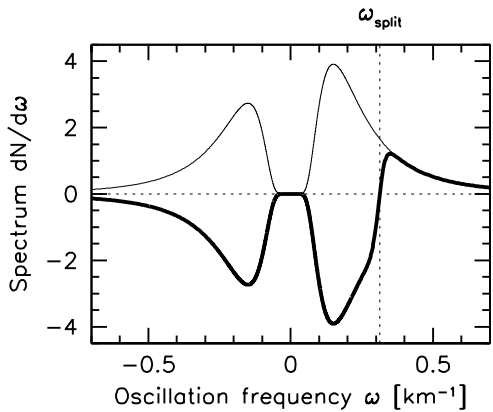}
\caption{Same as Fig.~\ref{fig:snspectrum}, now with $\omega=\Delta
m^2/2E$ as independent variable: $\omega<0$ is for
antineutrinos, $\omega>0$ for neutrinos.\label{fig:snspectrum2}}
\end{figure}

The split phenomenon becomes clearer when the spectra are represented
in terms of the oscillation frequency $\omega \equiv \Delta m^2/2E$
instead of the energy itself~\cite{Raffelt:2007cb}. In the evolution
equations the only difference between neutrinos and antineutrinos is
that the latter appear with a negative $\omega$. Therefore, one should
think of the neutrino and antineutrino spectra as a single continuum
$-\infty<\omega<+\infty$. The two spectra merge at the point
$\omega=0$ that represents both a neutrino or antineutrino of infinite
energy. In Fig.~\ref{fig:snspectrum2} we show the example of
Fig.~\ref{fig:snspectrum} in terms of the $\omega$--variable where
$\omega_{\rm split}=0.314~{\rm km}^{-1}$. All modes with
$\omega<\omega_{\rm split}$ are transformed whereas the ones with
$\omega>\omega_{\rm split}$ stay in their original flavor.

The split occurs among the neutrinos ($\omega_{\rm split}>0$) because
a SN core deleptonizes, producing an excess $\nu_e$ flux. On the other
hand, an excess of the $\bar\nu_e$--flux is typical for the
leptonizing accretion torus of merging neutron stars that are probably
the central engines of short gamma-ray bursts. Here we have
$\omega_{\rm split}<0$ and the situation is reversed: All modes with
$\omega>\omega_{\rm split}$ convert (Sec.~\ref{sec:splitfrequency}).
However, we will always use an excess $\nu_e$ flux without loss of
generality.

The interpretation of the split phenomenon proposed by Duan et
al.~\cite{Duan:2007mv} motivated us to develop a quantitative theory
of this effect based on the idea of adiabatic evolution of individual
modes~\cite{Raffelt:2007cb}. While the self-interacting neutrino
ensemble follows a set of nonlinear equations of motion (EOMs), the
evolution of individual modes can be written in the form of
single-particle EOMs where the neutrino-neutrino interaction is
included self-consistently by a slowly varying ``mean field'' in the
single-mode Hamiltonians. The crucial step was to obtain two sum rules
(integrals over the neutrino spectrum in the $\omega$--variable) that
had to equal certain conserved quantities, notably the flavor lepton
number of the system~\cite{Raffelt:2007cb}. These sum rules provide an
explicit adiabatic solution of the problem.

The degree of adiabaticity manifests itself in the ``sharpness'' of
the spectral split. The adiabatic evolution is driven by the
decreasing neutrino-neutrino interaction strength as a function of
distance from the SN core. As we dial the scale height to larger
values, the spectral split becomes sharper, approaching a step
function in the limit of an infinitely slow evolution.

This situation is similar to the usual MSW case, for example in the
outer layers of a SN envelope. The $\nu_e$ flux streaming through a
resonance region completely converts, for the normal hierarchy, while
the antineutrinos remain unchanged.  In other words, in the case of
perfect adiabaticity the usual MSW effect is a spectral-split
phenomenon with $\omega_{\rm split}=0$. In this sense the main effect
of the neutrino-neutrino interaction is to shift $\omega_{\rm split}$
to a non-vanishing value.  We will further discuss this analogy in
the context of adiabaticity breaking in Sec.~\ref{sec:transients}.

The main purpose of our paper is to elaborate several issues that were
kept short in our previous publication~\cite{Raffelt:2007cb}. In
addition, we will pay greater attention to the actual definition of
what we mean with adiabaticity and we will study phenomena caused by
deviations from a purely adiabatic evolution. We begin in
Sec.~\ref{sec:framework} with a description of the framework for our
study: the EOMs and our simplifications and approximations. In
Sec.~\ref{sec:adiabatic} we present the adiabatic solution in the form
of two sum rules. The following
Secs.~\ref{sec:neutrinos}--\ref{sec:twovectors} are dedicated to
explicit analytic and numerical solutions of the EOMs in particular
cases.  In Sec.~\ref{sec:transients} we study the impact of deviations
from perfect adiabaticity. We conclude in Sec.~\ref{sec:conclusion}.

\section{Framework and Simplifications}          \label{sec:framework}

\subsection{Equations of motion (EOMs)}

Nonlinear flavor transformations caused by neutrino-neutrino
interactions involve two main complications. One is the energy
dependence of the effect, a question that we will address here. The
other is the ``multi-angle'' nature of the transformations.  If a
neutrino gas is not perfectly isotropic, neutrinos moving in different
directions feel a different refractive effect caused by the other
neutrinos because the interaction energy between two neutrinos depends
on their relative direction as $(1-\cos\theta)$.  As a consequence,
one expects kinematical decoherence of the flavor content of different
angular modes. For equal densities of neutrinos and antineutrinos,
decoherence indeed occurs and is even self-accelerating in that an
infinitesimally small anisotropy is enough to trigger an exponential
runaway towards complete flavor equilibrium~\cite{Raffelt:2007yz}. On
the other hand, if the neutrino-neutrino interaction decreases slowly
and if the asymmetry between $\nu_e$ and $\bar\nu_e$ is sufficiently
large, one observes numerically that decoherence is suppressed. In the
SN context, the deleptonization flux appears to be large enough to
prevent multi-angle decoherence~\cite{EstebanPretel:2007ec}.

Therefore, we limit ourselves to an isotropic gas with a slowly
decreasing density. The ensemble is fully characterized by a matrix of
densities in flavor space for each spectral component $E$, separately
for neutrinos and antineutrinos. The energies enter only through the
vacuum oscillation frequencies $\omega \equiv \Delta m^2/2E$ which
determine the behavior of different modes. Therefore, it is more
convenient to use $\omega$ directly to label the modes. In the
two-flavor context we use the usual polarization vectors so that the
ensemble is represented by two sets of polarization vectors ${\bf
  P}_\omega$ and $\bar{\bf P}_\omega$. The global polarization
vectors,
\begin{equation}
 {\bf P}=\int_0^\infty d\omega\,{\bf P}_\omega
 \hbox{\quad and\quad}
 \bar {\bf P}=\int_0^\infty d\omega\,\bar{\bf P}_\omega,
\end{equation}
are initially normalized as
\begin{equation}
 P=|{\bf P}|=1
 \hbox{\quad and\quad}
 \bar P=|\bar {\bf P}|=\alpha<1\,.
\end{equation}
Notice that ${\bf P}_\omega$ and $\bar{\bf P}_\omega$ are densities
relative to the $\omega$ variable so that they have the dimension of
$\omega^{-1}$. The individual lengths $P_\omega=|{\bf P}_\omega|$ and
$\bar P_\omega=|\bar{\bf P}_\omega|$ are conserved as long as
oscillations are the only form of evolution, i.e., in the absence of
dynamical decoherence caused by collisions. In contrast, the total
lengths $P$ and $\bar P$ are only conserved when all individual
polarization vectors evolve in the same~way. $P_\omega$ and $\bar
P_\omega$ are the spectra of frequencies given by an initial
condition.

We assume that initially all ${\bf P}_\omega$ and $\bar{\bf P}_\omega$
are aligned. In other words, all neutrinos and antineutrinos begin in
the electron flavor. It is straightforward to generalize this
assumption to a situation where some modes begin in the $x$--flavor
(``spectral cross-over,'' see Sec.~\ref{sec:crossover}). In this more
general case, the ${\bf P}_\omega$ and $\bar{\bf P}_\omega$ may
initially also be anti-aligned relative to the common flavor
direction.

Following the notation of Ref.~\cite{Hannestad:2006nj} let us
introduce the difference vector
\begin{equation}
{\bf D} \equiv {\bf P}-\bar{\bf P},
\end{equation}
representing the net lepton number. Using the conventions of
Ref.~\cite{Sigl:1992fn}, the EOMs~are
\begin{eqnarray}\label{eq:eom1}
 \partial_t{\bf P}_\omega&=&\left(+\omega{\bf B} +\lambda{\bf L}
 +\mu {\bf D}\right)\times{\bf P}_\omega\,,
 \nonumber\\*
 \partial_t\bar{\bf P}_\omega&=&\left(-\omega{\bf B} +\lambda{\bf L}
 +\mu {\bf D} \right)
 \times\bar{\bf P}_\omega\,,
\end{eqnarray}
where ${\bf B} = (\sin 2 \theta, 0, \cos 2 \theta)$ is the mass
direction and $\theta$ the vacuum mixing angle. For the inverted mass
hierarchy $\theta \sim \pi/2$.  The unit vector ${\bf L} = (0, 0, 1)$
is the weak charged-current interaction direction so that ${\bf
  B}\cdot{\bf L}=\cos2\theta$. Finally, $\lambda\equiv\sqrt{2}G_F n_e$
is the usual matter potential and $\mu \equiv \sqrt{2}G_F n_\nu$ the
potential due to neutrino-neutrino interactions with $n_e$ and $n_\nu$
being the electron and neutrino densities. We use time $t$ instead of
radius $r$ because we study an isotropic system that evolves in time
instead of a single-angle system evolving in space.

In the SN context, the neutrino flux dilutes as $r^{-2}$.  Moreover,
the trajectories become more collinear with distance, leading to an
average suppression of the interaction energy by another factor
$r^{-2}$ \cite{Qian:1995ua} so that altogether we should use
$\mu(t)\propto t^{-4}$. However, in the adiabatic limit the exact form
of $\mu(t)$ is irrelevant as long as its rate of change is small.  For
our numerical examples it is more convenient to use the profile
\begin{equation}\label{eq:muexp}
\mu(t)=\mu_0\,\exp(-t/\tau)\,.
\end{equation}
The exponential form is simpler because it involves the same scale
$|d\ln \mu(t)/dt|^{-1}=\tau$ at all times. Adjusting $\tau$ allows one
to control the adiabaticity of the transition with a single
parameter. Assuming the atmospheric $\Delta m^2$ and typical SN
neutrino energies, we find that $\omega_0=0.3~{\rm km}^{-1}$ is
typical for the average vacuum oscillation frequency. The crucial
action takes place at the relatively large radius of a few hundred
kilometers. Together with the $r^{-4}$ scaling this implies a
characteristic rate of change of $0.01~{\rm km}^{-1}$. Therefore, in
Eq.~(\ref{eq:muexp}) a value $\tau=(0.03\,\omega_0)^{-1}$ roughly
captures the situation in a~SN.

\subsection{Removing the usual matter effect}

Unless there is an MSW resonance in the dense-neu\-trino region, one
can eliminate $\lambda{\bf L}$ from Eq.~(\ref{eq:eom1}) by going into
a rotating frame~\cite{Duan:2005cp, Hannestad:2006nj}. The price for
this transformation is that in the new frame ${\bf B}$ has a static
component $B_L=B\cos2\theta$ along the ${\bf L}$ direction and a
fast-rotating one in the transverse direction. Assuming that the
effect of the fast-rotating component averages to zero, we may
consider the equivalent situation where the usual matter is completely
absent and initially all polarization vectors are aligned with ${\bf
  B}$ (which really is the projection of the true ${\bf B}$ on the
weak-interaction direction ${\bf L}$). This is equivalent to a
vanishing vacuum mixing angle.

Of course, this is not the complete story because for a strictly
vanishing vacuum mixing angle, no transformation effects would take
place at all. The rotating component of ${\bf B}$ disturbs the perfect
alignment of ${\bf P}_\omega$ and $\bar{\bf P}_\omega$, providing a
kick start for flavor conversions~\cite{Duan:2005cp}. For a system
consisting of equal numbers of neutrinos and antineutrinos, the
equations were solved analytically~\cite{Hannestad:2006nj}. The net
effect was found to be equivalent to a small (but nonvanishing)
effective mixing angle, although the evolution of the polarization
vectors is initially modulated by the rotating ${\bf B}$ field with
the frequency~$\lambda$. We have here a different system where the
density of neutrinos exceeds that of antineutrinos and where the
evolution of the system is driven by the $\mu(t)$ variation.
Numerically one finds that once more the effect of matter is
equivalent to reducing the vacuum mixing angle to a certain effective
mixing angle $\theta_{\rm eff}$ similar to the usual in-medium mixing
angle~\cite{EstebanPretel:2007ec}, but an analytic treatment of how
ordinary matter modifies the initial evolution is not available.

Motivated by this evidence we make the simplification of ignoring the
ordinary matter term entirely, schematically accounting for it by a
small initial effective mixing angle. Notice that in the absence of
the interaction vector ${\bf L}$ we define the effective mixing angle
as the angle between ${\bf D}$ and ${\bf B}$. Our treatment is exact
for a system where the only interaction is that of the neutrinos among
each other. While we cannot claim with mathematical rigor that this
system is a faithful proxy for what happens in a real SN, numerical
observations suggest that this is the case.

\subsection{EOMs in the {\boldmath$\omega$} variable}

The equation of motion for $\bar{\bf P}_\omega$ differs from that for
${\bf P}_\omega$ only by the sign of frequency: $\omega\to-\omega$.
So, the only difference for antineutrinos is that in vacuum they
oscillate ``the other way round.'' Therefore, instead of using
$\bar{\bf P}_\omega$ we may extend ${\bf P}_\omega$ to negative
frequencies such that $\bar{\bf P}_{\omega} = {\bf P}_{-\omega}$
($\omega > 0$) and use only ${\bf P}_\omega$ with
$-\infty<\omega<+\infty$. After eliminating the matter term
$\lambda{\bf L}$ the EOMs then take on the simple form
\begin{equation}\label{eq:eom1a}
 \dot{\bf P}_\omega=\left(\omega{\bf B}
 +\mu {\bf D}\right)\times{\bf P}_\omega\,.
\end{equation}
A compact expression for the difference vector is
\begin{equation}
\label{def-d}
{\bf D}= \int_{-\infty}^{+\infty}d\omega\,s_\omega\,{\bf P}_\omega,
\end{equation}
where
\begin{equation}
s_\omega \equiv {\rm sign}(\omega)= \frac{\omega}{|\omega|}\,.
\end{equation}
Integrating both sides of Eq.~(\ref{eq:eom1a}) over $s_\omega
d\omega$ provides
\begin{equation}\label{eq:eomD}
 \dot{\bf D} = {\bf B} \times{\bf M}\,,
\end{equation}
where
\begin{equation}
{\bf M} \equiv
 \int_{-\infty}^{+\infty} d\omega\, s_\omega \omega\,
 {\bf P}_\omega
\end{equation}
is the ``effective magnetic moment'' of the system.

The EOM for ${\bf D}$ is not a closed differential equation. However,
when $\mu$ is large, all ${\bf P}_\omega$ ($-\infty<\omega<\infty$)
remain pinned to each other, and therefore ${\bf M} \propto {\bf D}$.
In this case, according to  Eq.~(\ref{eq:eomD}) the  collective
vector ${\bf D}$ precesses around ${\bf B}$ with the synchronization
frequency~\cite{Pastor:2001iu}
\begin{equation}\label{eq:wsynch}
 \omega_{\rm synch}=\frac{M}{D}=
 \frac{\int_{-\infty}^{+\infty} d\omega\,s_\omega\omega\,P_\omega}
 {\int_{-\infty}^{+\infty} d\omega\,P_\omega}\,.
\end{equation}
If $\mu$ is suddenly turned off, all ${\bf P}_\omega$ henceforth
precess with their individual $\omega$. The transverse component of
${\bf D}$ quickly averages to zero (kinematical decoherence), whereas
the component along ${\bf B}$ is conserved. We will here study the
opposite limit where $\mu$ decreases slowly, leading to a very
different final result.

\subsection{Lepton-number conservation}

Since ${\bf B}$ is constant, Eq.~(\ref{eq:eomD}) shows that
$\partial_t({\bf D}\cdot{\bf B})=0$ and therefore
\begin{equation}\label{eq:DB}
D_\parallel = {\bf B} \cdot {\bf D} = {\rm const.},
\end{equation}
where here and henceforth we will denote the component of a vector
parallel to ${\bf B}$ with the index $\parallel$, the transverse
component with the index $\perp$.

The physical interpretation of the conservation law Eq.~(\ref{eq:DB})
is what we call ``flavor lepton-number conservation'' or simply
``lepton-number conservation.'' Rigorously it only means that in the
absence of matter the projection of ${\bf D}$ on ${\bf B}$ is
conserved. However, in all examples of practical interest in the SN
context, the effective mixing angle is small so that initially ${\bf
D}$ and ${\bf B}$ are nearly collinear. Therefore, the net $\nu_e$
flux from deleptonization is approximately conserved until an MSW
resonance is encountered. Collective effects only induce pair
transformations of the form $\nu_e\bar\nu_e\to\nu_x\bar\nu_x$,
whereas the excess of the $\nu_e$ flux from deleptonization is
conserved.

\section{Adiabatic solution}
\label{sec:adiabatic}

\subsection{Adiabaticity and co-rotating plane}

To understand the evolution of our system in the  limit of a slowly
changing $\mu$, we rewrite the EOMs as
\begin{equation}
\label{eq:eom2}
 \dot{\bf P}_\omega={\bf H}_\omega\times{\bf P}_\omega\,,
\end{equation}
where $-\infty<\omega<+\infty$. We have introduced an ``individual
Hamiltonian'' for each mode
\begin{equation}
\label{ind-ham}
{\bf H}_\omega = \omega{\bf B} +\mu {\bf D}\,.
\end{equation}

According to the standard notion, the evolution is adiabatic when the
rate of change $\omega^{\rm eff}({\bf H}_\omega)$ of each  ${\bf
H}_\omega$ is slow compared to the precession speed $\omega^{\rm
eff}({\bf P}_\omega)$ of ${\bf P}_\omega$:
\begin{equation}
\omega^{\rm eff}({\bf P}_\omega) \gg
\omega^{\rm eff}({\bf H}_\omega)\,.
\label{ad-non}
\end{equation}
In this case ${\bf P}_\omega$ follows ${\bf H}_\omega$. In general
${\bf P}_\omega$ need not coincide with ${\bf H}_\omega$ but rather
moves around ${\bf H}_\omega$ on the surface of a cone whose axis
coincides with ${\bf H}_\omega$ and the cone angle is constant.

Let us check the adiabaticity condition for our system. Since $|D|
\leq 1 + \alpha$, the speed (effective frequency) of the ${\bf
P}_\omega$ motion can be estimated according to Eq.~(\ref{eq:eom2})
as
\begin{equation}
\omega^{\rm eff}({\bf P}_\omega)
\leq \sqrt{\omega^2 + \mu^2 (1 + \alpha)^2 }\,.
\end{equation}
${\bf D}$ and therefore ${\bf H}_\omega$ precess around ${\bf B}$
with the synchronization frequency of Eq.~(\ref{eq:wsynch}):
\begin{equation}
\omega^{\rm eff}({\bf H}_\omega) = \omega_{\rm synch} \sim \omega_0,
\end{equation}
where $\omega_0$ is some typical frequency in the spectrum. The
Hamiltonian ${\bf H}_\omega$ changes both because of the
$\mu$--decrease and because of the precession of ${\bf D}$. For $\mu
\sim \omega_0$ or smaller we have
\begin{equation}
\omega^{\rm eff}({\bf P}_\omega) \sim
\omega^{\rm eff}({\bf H}_\omega),
\label{ad-non2}
\end{equation}
and therefore the adiabaticy condition that requires much faster
motion of ${\bf P}_\omega$ is not fulfilled. Apparently though the
adiabaticity condition is fulfilled for  $\mu \gg \omega_0$.

However, it is usually not stressed that the question of adiabaticity
depends on the coordinate system and therefore adiabaticity has a
relative meaning.  For a single polarization vector the evolution is
trivially adiabatic in a reference frame co-moving with its evolution,
although this observation is useless as a means for solving the
EOM. It is also trivial to find a frame where the evolution is
non-adiabatic. Even for the case of the ordinary MSW effect we can go
into a frame that rotates with a large frequency around the
$z$--direction. In this frame the evolution does not look
adiabatic. Of course, the final outcome is unchanged because the
physical evolution does not depend on the coordinate system in which
it is discussed, but the interpretation is different. In the
fast-rotating frame the ordinary MSW effect is interpreted as a
parametric resonance caused by a ${\bf B}$--field with a fast-rotating
transverse component.

Let us return to the case at hand. Inspecting Eq.~(\ref{ind-ham})
reveals that the fast motion of the single-mode Hamiltonians in the
``laboratory frame'' is caused by the fast motion of the single common
vector ${\bf D}$. If the motion of ${\bf D}$ is essentially a
precession around ${\bf B}$ as suggested by Eq.~(\ref{eq:eomD}), the
relevant co-moving frame is simply the plane of ${\bf B}$ and ${\bf
  D}$ that rotates around ${\bf B}$. In other words, while both ${\bf
  H}_\omega$ and ${\bf P}_\omega$ precess fast around ${\bf B}$, in
the co-rotating frame they may move slowly relative to each other and
relative to the common frame which then plays the role of the
adiabatic frame for all ${\bf P}_\omega$.

Going to a rotating frame to simplify the EOMs was first suggested by
Duan et~al.~\cite{Duan:2005cp}, but later they have questioned that
the evolution can then be called adiabatic because it is not
adiabatic in the laboratory frame~\cite{Duan:2007fw}. We believe that
the notion of adiabaticity is nevertheless appropriate because the
laboratory frame is not special. (In fact, we have already
transformed away the usual matter effect by going into a rotating
frame which for us plays the role of the laboratory frame.)
The evolution of the multi-mode system should be called
adiabatic if there exists a common frame in which all modes evolve
adiabatically.

In the present context, the co-rotating frame spanned by ${\bf B}$
and ${\bf D}$ naturally provides a common adiabatic frame.
Independently of the motion of the individual ${\bf P}_\omega$ that
define the overall ${\bf D}$, the single-mode Hamiltonians ${\bf
H}_\omega$  always lie in a single plane spanned by the vectors ${\bf
B}$ and ${\bf D}$. Relative to the laboratory frame, this common or
co-rotating plane moves around ${\bf B}$ with the instantaneous
co-rotating frequency $\omega_{\rm c}$. The Hamiltonians in the
co-rotating frame are
\begin{equation}\label{eq:Hi}
 {\bf H}_\omega=(\omega-\omega_{\rm c})\,{\bf B}
 +\mu {\bf D}\,.
\end{equation}
We use the same notation in both frames because the relevant
components $H_{\omega\parallel}$, $H_{\omega \perp}$,
$D_{\parallel}$, and $D_{\perp}$ remain invariant under transition
from the one frame to another.

Let us find the co-rotating frequency $\omega_{\rm c}$. By assumption
all ${\bf P}_\omega$, and consequently ${\bf M}$, stay in the
co-rotating plane. Therefore we can decompose
\begin{equation}\label{decompose}
 {\bf M} = b\, {\bf B} + \omega_{\rm c} {\bf D}
\end{equation}
and rewrite the EOM of Eq.~(\ref{eq:eomD}) as
\begin{equation}
\label{eq:eomD2}
\partial_t{\bf D} =  \omega_{\rm c} {\bf B} \times{\bf D}.
\end{equation}
Projecting Eq.~(\ref{decompose}) on the transverse plane we find
$\omega_{\rm c} = M_{\perp} / D_\perp$ or explicitly
\begin{equation}\label{eq:Dfreq}
 \omega_{\rm c}=
 \frac{\int_{-\infty}^{+\infty} d\omega\,s_\omega\,\omega\,P_{\omega\perp}}
 {\int_{-\infty}^{+\infty} d\omega\,s_\omega P_{\omega\perp}}
 =\frac{\int_{-\infty}^{+\infty} d\omega\,s_\omega\,\omega\,P_{\omega\perp}}
 {D_\perp}\,.
\end{equation}
When $\mu\to\infty$ and all ${\bf P}_\omega$ are aligned, this is
identical with the synchronization frequency Eq.~(\ref{eq:wsynch}).

\subsection{Adiabatic solution in terms of sum rules}

Initially when $\mu$ is very large, all individual Hamiltonians are
essentially aligned with ${\bf D}$. In turn, ${\bf D}$ is aligned with
the weak-interaction direction if initially all polarization vectors
${\bf P}_\omega$ are aligned with that direction. In other words, all
neutrinos are prepared in interaction eigenstates and initially ${\bf
  P}_\omega \propto {\bf H}_\omega$. The adiabatic evolution would
imply that if $\mu$ changes slowly enough, ${\bf P}_\omega$ follows
${\bf H}_\omega (\mu)$ and therefore remains aligned with ${\bf
  H}_\omega (\mu)$ at later times as well. So the adiabatic solution
of the EOMs for our initial condition is given by
\begin{equation}
\label{eq:hialigned}
{\bf P}_\omega (\mu) = \hat{\bf H}_\omega (\mu) \,P_\omega\,,
\end{equation}
where $P_\omega=|{\bf P}_\omega|$ and $\hat{\bf H}_\omega \equiv {\bf
H}_\omega/|{\bf H}_\omega|$ is a unit vector in the direction of the
Hamiltonian.

According to the solution Eq.~(\ref{eq:hialigned}) all ${\bf
P}_\omega$ being confined to the co-rotating plane evolve in this
frame according to the change of $\mu$. The solution is implicit
since $\hat{\bf H}_\omega$ depends via  ${\bf D} = {\bf D}({\bf
P}_\omega)$ on the polarization vectors:
\begin{equation}
{\bf H}_\omega  = {\bf H}_\omega ({\bf P}_\omega), \qquad
{\bf P}_\omega = {\bf P}_\omega (\hat{\bf H}_\omega)\, .
\end{equation}
We will see that there is a consistent solution which satisfies these
relations. Notice that by imposing the ``alignment'' condition
Eq.~(\ref{eq:hialigned}) on ${\bf P}_\omega$, we also restrict the
Hamiltonian, since ${\bf H}_\omega$ depends on ${\bf P}_\omega$.

Let us find explicit solutions $P_{\omega\parallel}$  and
$P_{\omega\perp}$. Since $P_{\omega}$ is conserved and simply the
spectrum of neutrinos given by the initial condition, we only need to
find the Hamiltonians Eq.~(\ref{eq:Hi}). They, in turn, are
completely determined by $\omega_{\rm c}(\mu)$ and $D_\perp(\mu)$
because $D_{\omega\parallel}$ is known to be constant as determined
by the lepton number. In other words, we need to find self-consistent
results for $\omega_{\rm c}(\mu)$ and $D_\perp(\mu)$, the component
transverse to ${\bf B}$, for a given $D_\parallel$.

The implicit adiabatic solution Eq.~(\ref{eq:hialigned}) gives us two
equations for each $\omega$ which we use to find $\omega_{\rm
c}(\mu)$ and $D_\perp(\mu)$. Projecting Eq.~(\ref{eq:hialigned}) onto
the perpendicular and parallel directions with respect to ${\bf B}$
provides
\begin{eqnarray}
P_{\omega\perp}&=&\frac{H_{\omega\perp}}{H_\omega} P_\omega,
\nonumber\\*
P_{\omega \parallel}&=&\frac{H_{\omega\parallel}}{H_\omega} P_\omega.
\end{eqnarray}
From Eq.~(\ref{eq:Hi}) we infer
\begin{eqnarray}
H_{\omega\perp}&=&\mu D_\perp,
\nonumber\\*
H_{\omega\parallel}&=&\omega-\omega_{\rm c}+\mu D_\parallel,
\end{eqnarray}
so that
\begin{eqnarray}
 P_{\omega\parallel}&=&
 \frac{(\omega-\omega_{\rm c}+\mu D_\parallel)\,P_\omega}
 {\sqrt{(\omega-\omega_{\rm c}+\mu D_\parallel)^2+(\mu D_\perp)^2}}\,,
 \label{eq:pz}\\*
 P_{\omega\perp}&=&\frac{\mu D_\perp\,P_\omega}
 {\sqrt{(\omega-\omega_{\rm c}+\mu D_\parallel)^2+(\mu D_\perp)^2}}\,.
 \label{eq:pperp}
\end{eqnarray}
Integration of these equations over $s_\omega d\omega$ gives us
\begin{eqnarray}
\label{eq:master2}
 \kern-1.5em
 D_\parallel&=& \int_{-\infty}^{+\infty}d\omega\,s_\omega\,
 \frac{(\omega-\omega_{\rm c}+\mu D_\parallel)\,P_\omega}
 {\sqrt{(\omega-\omega_{\rm c}+\mu D_\parallel)^2+(\mu D_\perp)^2}}\,,\\*
 \label{eq:master1}
 \kern-1.5em
 1&=&\int_{-\infty}^{+\infty}d\omega\,s_\omega\,\frac{P_\omega}
 {\sqrt{[(\omega-\omega_{\rm c})/\mu+D_\parallel]^2+D_\perp^2}}\,.
\end{eqnarray}
For a given $D_\parallel$ and  spectrum $P_\omega$, we can determine
$\omega_{\rm c}$ and $D_\perp$ from Eqs.~(\ref{eq:master2})
and~(\ref{eq:master1}) for any $\mu$ and thus find explicit
solutions.

Equations~(\ref{eq:master2}) and~(\ref{eq:master1}) can be considered
as sum rules---integrals over frequencies that should be equal to
certain conserved numbers. They can be added to each other with
various factors so that one can rewrite them in different forms
depending on convenience.  For example, Eq.~(\ref{eq:master2}) can be
rewritten with the use of Eq.~(\ref{eq:master1}) as
\begin{equation}\label{eq:master2aa}
 \omega_{\rm c}=\int_{-\infty}^{+\infty} d\omega\,s_\omega\,
 \frac{\omega\,P_\omega}
 {\sqrt{[(\omega-\omega_{\rm c})/\mu+D_\parallel]^2+D_\perp^2}}
\end{equation}
which coincides with the form derived in our previous paper
\cite{Raffelt:2007cb}.

While different forms of the sum rules are equivalent, their limiting
behavior for $\mu\to0$ or $\mu\to\infty$ is not always equally
manifest. For example, the physically intuitive expression
Eq.~(\ref{eq:Dfreq}) for $\omega_{\rm c}$ is valid only for
$D_\perp(\mu) \neq 0$ while for $D_\perp = 0$ it has a 0/0 feature.
On the other hand, Eq.~(\ref{eq:master2aa}) for $\mu\to\infty$ is
equivalent to the synchronization frequency Eq.~(\ref{eq:wsynch})
without any ambiguity.

Inserting the solution  Eq.~(\ref{eq:hialigned}) into the EOM
Eq.~(\ref{eq:eom2}) we obtain
\begin{equation}
\label{eq:adeq}
 \partial_t{\bf H}_\omega (\mu) = 0
\end{equation}
which is satisfied exactly if $\partial_t \mu = 0$  and
$\partial_t{\bf D} = 0$. That is, the medium has constant neutrino
density and the difference vector does not change. The latter is
satisfied when all polarization vectors stay unchanged, but this may
not be the only possibility. We will call this self-consistent
solution the ``static solution.'' For a slowly varying density the
condition Eq.~(\ref{eq:adeq}) is only approximately satisfied. As in
the usual MSW case, Eq.~(\ref{eq:hialigned}) provides an approximate
solution of the EOMs, when the dependence of the Hamiltonian on time
is negligible. The adiabatic solution (the lowest order term in the
adiabatic perturbation theory) is given by a continuous set of static
(instantaneous) solutions.

Let us comment on the static solution. It consists of all ${\bf
P}_\omega$ being confined to the co-rotating plane. The motion of all
polarization vectors is limited to a common precession around ${\bf
B}$ with frequency $\omega_{\rm c}$, each of them having a different
zenith angle relative to ${\bf B}$. Duan et~al. have termed this form
of motion the ``pure precession mode''~\cite{Duan:2007mv}. In
contrast to the usual MSW case, the static solution implies an
additional condition ${\bf D} = {\rm const}$. which does not mean in
general that the individual polarization vectors are static. Another
issue  is the stability of the solution since ${\bf D}$ is a
dynamical system and not a rigid vector. Furthermore, separate
equalities $\partial_t \mu = 0$  and $\partial_t{\bf D}= 0$ may not
be the only solutions of Eq.~(\ref{eq:adeq}). Our numerical studies
show, however, that the static solutions are not only
self-consistent, but also stable.

For every spectrum $P_\omega$ the pure precession solutions
are characterized by the two parameters $D_\parallel$ and $\mu$. For
$\mu\to\infty$ all Hamiltonians are aligned and thus all ${\bf
  P}_\omega$ are aligned as well, provided that  they are in a pure
precession mode. Assuming the neutrinos are produced in interaction
eigenstates, all ${\bf P}_\omega$ are tilted relative to ${\bf B}$
with twice the effective mixing angle so that
\begin{equation}
D_\parallel=\cos 2\theta_{\rm eff}\,\int_{-\infty}^{+\infty}
d\omega\,s_\omega\,P_\omega\,.
\end{equation}
In this sense we have, for every spectrum $P_\omega$, a two-parameter
family of static solutions given in terms of $\cos 2\theta_{\rm eff}$
and $\mu$ or equivalently, in terms of $D_\parallel$ and~$\mu$.

\subsection{Split frequency}

\label{sec:splitfrequency}

In the limit $\mu\to\infty$ all polarization vectors are aligned with
each other in a direction given by the choice of $\cos 2\theta_{\rm
  eff}^{\infty}$, i.e., by the initial condition.  In the opposite
limit, $\mu\to 0$, the solution is given by
\begin{equation}\label{eq:Hiad}
 {\bf H}_\omega  \rightarrow  (\omega-\omega_{\rm c}^0)\,{\bf B}\,,
\end{equation}
where $\omega_{\rm c}^0\equiv\omega_{\rm c}(\mu\to0)$.  All
Hamiltonians and thus all ${\bf P}_\omega$ with $\omega>\omega_{\rm
  c}$ are aligned with ${\bf B}$, whereas those with
$\omega<\omega_{\rm c}^0$ are anti-aligned. Therefore, we have a
spectral split at the frequency
\begin{equation}
\omega_{\rm split}=\omega_{\rm c}^0
\end{equation}
which usually is not equal to zero.

$D_\perp=0$ because in the end all polarization vectors
are (anti)aligned with ${\bf B}$. Moreover, for $\mu\to0$
Eq.~(\ref{eq:master2}) approaches a well-defined limit,
\begin{eqnarray}
\label{lcharge}
D_\parallel &=& \int_{-\infty}^{+\infty}d\omega\,s_\omega\,
 \frac{(\omega-\omega_{\rm split})}
 {\sqrt{(\omega-\omega_{\rm split})^2 }}\,P_\omega
 \nonumber\\*
 &=&\int_{-\infty}^{+\infty}d\omega\,s_\omega\,
 s_{\omega-\omega_{\rm split}}\,P_\omega\,.
\end{eqnarray}
For our usual case of an excess neutrino flux ($D_\parallel>0$ and
$\omega_{\rm split}>0$) this equation is identical with
\begin{equation}
\label{eq:wsplit}
 D_\parallel=\int_{-\infty}^0 P_\omega\,d\omega
 -\int_{0}^{\omega_{\rm split}} P_\omega\,d\omega
 +\int_{\omega_{\rm split}}^{+\infty} P_\omega\,d\omega\,.
\end{equation}
The first term enters with a positive sign because antineutrinos
enter lepton number with a negative sign, but are now
oriented opposite to ${\bf B}$, the next term has a negative sign
because it represents those neutrino modes that in the end are
oriented against ${\bf B}$, and the third term has a positive sign
because it represents neutrinos with orientation along ${\bf B}$.

In practice we are mostly interested in the application to SN physics
where the effective mixing angle is small. In this case there are
other intuitive ways to state the condition of lepton-number
conservation and to determine the split frequency. We can rewrite the
definition of ${\bf D}$ in Eq. (\ref{def-d}) for $\theta_{\rm
  eff}^{\infty} \approx 0$ in the following explicit form
\begin{equation}\label{eq:wsplit2}
 D_\parallel \approx -\int_{-\infty}^0 P_\omega\,d\omega
 +\int_{0}^{\omega_{\rm split}} P_\omega\,d\omega
 +\int_{\omega_{\rm split}}^{+\infty} P_\omega\,d\omega\,.
\end{equation}
Now we have two equations, Eq.~(\ref{eq:wsplit}) and
Eq.~(\ref{eq:wsplit2}), for integrals over different parts of spectra.
The first equation gives us the total lepton number of the initial
state where antineutrinos enter with a negative sign, neutrinos with
a positive sign. The second equation simply reflects that all modes
below $\omega_{\rm split}$ were flipped.

Subtracting or adding these equations eliminates different parts of
the spectrum and reveals
\begin{equation}
\label{equa2}
 \int_{- \infty}^{0}d\omega\, P_\omega
 = \int_{0}^{\omega_{\rm split}} d\omega\, P_\omega
\end{equation}
and
\begin{equation}
\label{equa3}
 D_\parallel =
 \int_{\omega_{\rm split}}^{+ \infty} d\omega\, P_\omega\,.
\end{equation}
Equation~(\ref{equa2}) has the interpretation that all antineutrinos
($-\infty<\omega< 0$) are converted and therefore to conserve lepton
number the same number of neutrinos should be converted. From the
continuity in $\omega$ we infer that those are neutrinos with
frequencies with $0<\omega<\omega_{\rm split}$.
Equation~(\ref{equa3}) has the interpretation that the net lepton
charge of the system should be determined by the high frequency part
of the neutrino spectrum.

For the case of a dominant antineutrino flux ($\alpha > 1$) we have
$D_\parallel < 0$, the split frequency is negative and the spectral
split occurs in the antineutrino channel. Now all modes with
frequencies above $\omega_{\rm split}$ are converted whereas the
modes with negative $\omega$ below $\omega_{\rm split}$ stay
unchanged. The sum rule for the determination of $\omega_{\rm split}$
now becomes
\begin{equation}
\label{lcharge-neg}
D_\parallel = - \int_{- \infty}^{\omega_{\rm split}} d\omega\, P_\omega
+ \int^{0}_{\omega_{\rm split}} d\omega\, P_\omega
- \int_{\omega_{0}}^{+ \infty} d\omega\, P_\omega\,.
\end{equation}
Instead of Eq.~(\ref{equa3}) we get
\begin{equation}
\label{equa3-neg}
D_\parallel =  - \int^{\omega_{\rm split}}_{\infty} d\omega\, P_\omega\,.
\end{equation}
The net charge now comes from negative frequencies. So in this case
the entire neutrino flux and high energy part of the antineutrino
flux change flavor whereas the low energy antineutrino flux is
unchanged.

\section{Solutions for Neutrinos}                \label{sec:neutrinos}

\subsection{Box spectrum}

As a first example for explicit adiabatic solutions we consider the
simplest possible case where neutrino-neutrino interactions lead to a
spectral split in the adiabatic limit: An ensemble consisting of
neutrinos only ($\alpha=0$), no antineutrinos and  no ordinary matter.
Now ${\bf D} = {\bf P}$ and correspondingly $D_\parallel =
P_\parallel$ and $D_\perp = P_\perp$. Of course, this is an abstract
example meant to illustrate the general phenomenon. In subsequent
sections we will turn to realistic situations relevant in a SN.

We use a flat distribution of oscillation frequencies with the
average $\omega_0$, i.e., the box-like spectrum
\begin{equation}\label{eq:boxspetrum}
 P_\omega=
 \cases{(2\omega_0)^{-1}&
 for~$0\leq\omega\leq2\omega_0$,\cr
 0&otherwise.\cr}
\end{equation}
With $P_\parallel$ the conserved projection of ${\bf P}$ on the ${\bf
  B}$--direction we find from Eqs.~(\ref{eq:master2aa})
and~(\ref{eq:wsplit}), recalling that Eqs.~(\ref{equa2})
  and~(\ref{equa3}) are valid for $\theta_{\rm eff}^{\infty}\approx
  0$,
\begin{equation}
\label{eq:boxlimits}
 \omega_{\rm c}=\omega_0\times
 \cases{1&for~$\mu\to\infty$,\cr
 (1-P_\parallel)&for~$\mu=0$.\cr}
\end{equation}
Notice that for $P_\parallel = 0$, corresponding to maximal initial
mixing, the frequency $\omega_{\rm c}=\omega_0$ is constant for the
entire evolution. For $P_\parallel = 1$ (vanishing mixing angle) we
have $\omega_{\rm c}= 0$, corresponding to the absence of any flavor
evolution. We will see in Sec.~\ref{sec:both} that in the presence of
antineutrinos even a very small mixing angle leads to a large effect,
whereas here we need to assume a non-zero mixing angle to obtain
something visible.

\begin{figure}
\includegraphics[width=0.75\columnwidth]{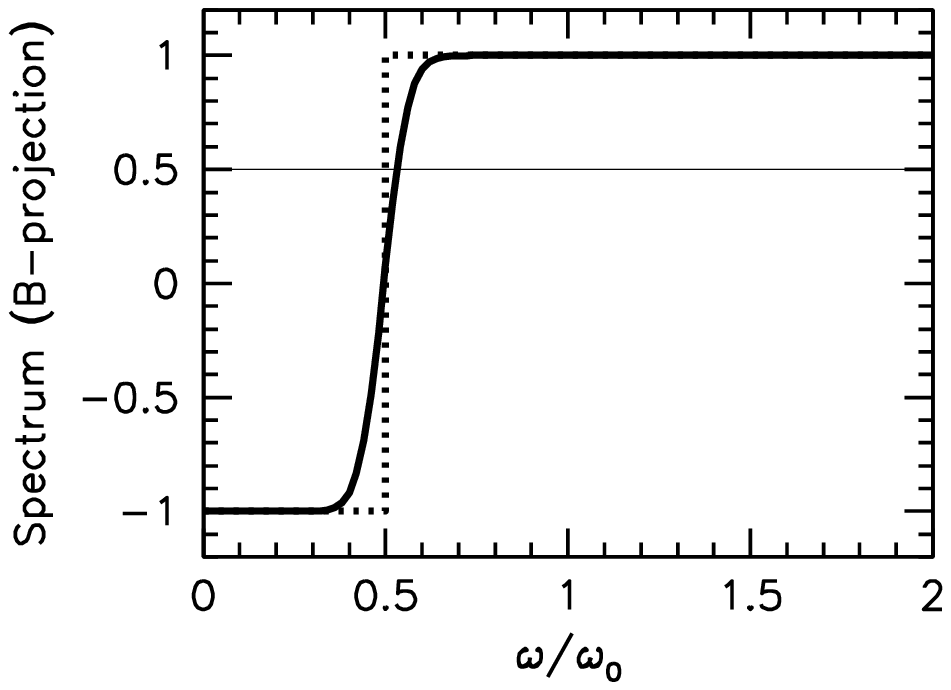}
\caption{Initial (thin line) and final (thick line) neutrino spectra
for the example of Eq.~(\ref{eq:boxspetrum}). Dotted: fully
adiabatic. Solid: numerical example with
$\tau=(0.03\,\omega_0)^{-1}$ in Eq.~(\ref{eq:muexp}).
\label{fig:boxspectrum}}
\vskip18pt
\includegraphics[width=0.8\columnwidth]{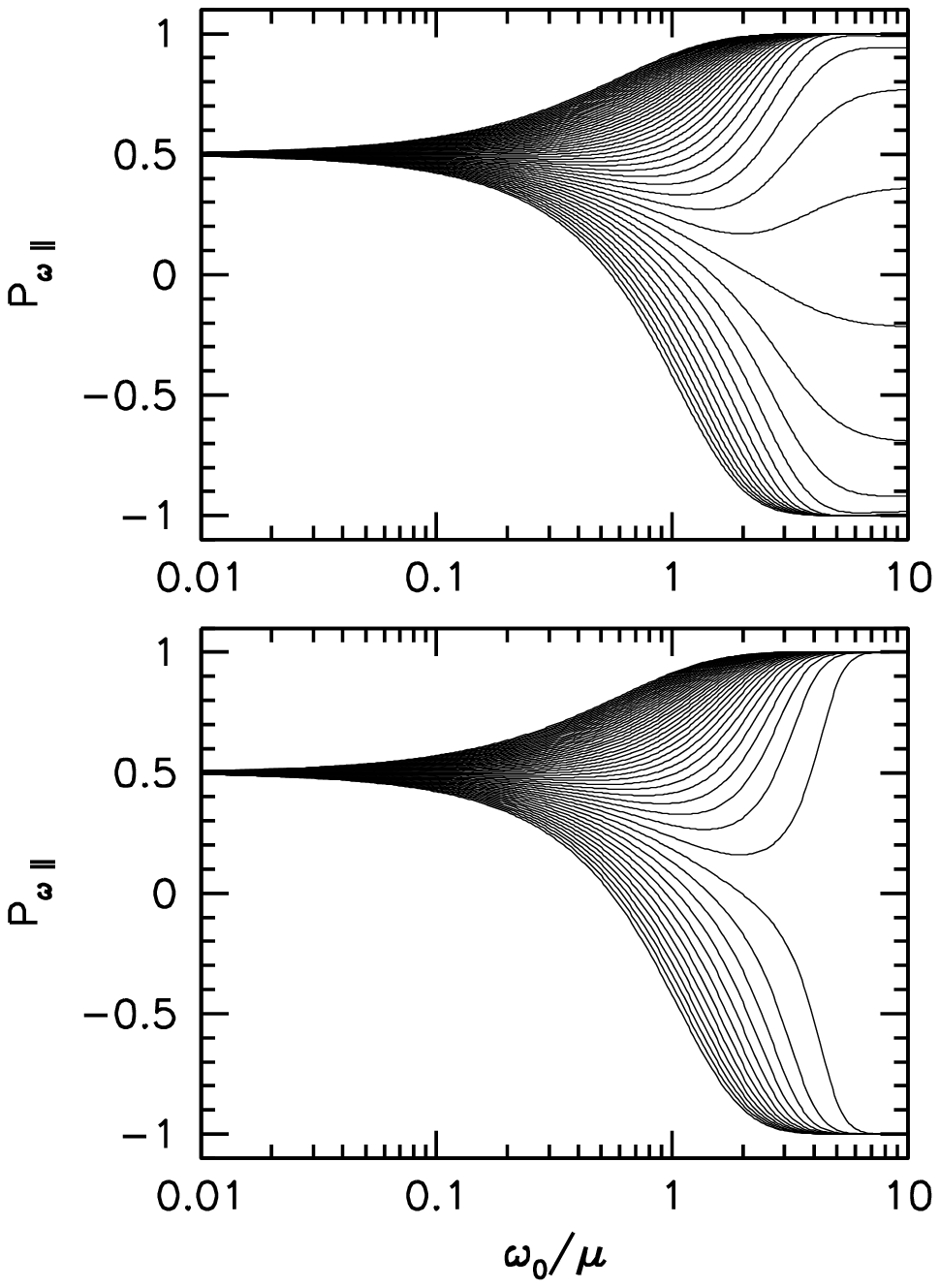}
\caption{Evolution of 51~equally spaced modes for the same example
as in Fig.~\ref{fig:boxspectrum}. {\it Top}: Numerical solution with
$\tau=(0.03\,\omega_0)^{-1}$. {\it Bottom}: Adiabatic solution
derived in Sec.~\ref{sec:boxsolution}.\label{fig:spectrumevolution}}
\end{figure}

In Fig.~\ref{fig:boxspectrum} we show the initial and final
distribution $P_{\omega\parallel}$ for the example
$P_\parallel=0.5$, so that according to Eq.~(\ref{eq:boxlimits})
$\omega_{\rm split}=\omega_{\rm c}^0 = 0.5\,\omega_0$. The dotted
line indicates the final distribution in the adiabatic limit, the
solid line is the result of a numerical example with $\mu(t)$ from
Eq.~(\ref{eq:muexp}) with $\tau=(0.03\,\omega_0)^{-1}$.
The spectral split indeed becomes sharper
with increasing $\tau$ (increasing adiabaticity) as discussed in more
detail in Sec.~\ref{sec:transients}.

We may also follow the evolution of the individual polarization
vectors. In Fig.~\ref{fig:spectrumevolution} we show
$P_{\omega\parallel}$ for 51 equally spaced modes as a function
of $\omega_0/\mu$ for the example of Fig.~\ref{fig:boxspectrum}. At
first, all of them begin with the common starting value
$P_{\omega\parallel} =  0.5$. Later they spread out and eventually
split, some of them approaching $+1$ and the others $-1$ as
predicted. Some of them first move down and then turn around as a
result of $\omega_{\rm c}$ changing as a function of time. As a
result of imperfect adiabaticity, a few modes do not reach $\pm1$
but get frozen before this destination.

\subsection{Analytic solution}
\label{sec:boxsolution}

In the adiabatic limit, the same results follow from our explicit
solution of the EOMs. For the box spectrum Eq.~(\ref{eq:boxspetrum})
the integrals Eqs.~(\ref{eq:master2}) and~(\ref{eq:master1}) are
easily performed. After some transformations we find
\begin{eqnarray}\label{eq:exact1}
 \frac{\omega_{\rm c}}{\omega_0}&=&1+
 P_\parallel\left(\frac{1}{\kappa} -
 \frac{e^{\kappa}+e^{-\kappa}}{e^{\kappa}-e^{-\kappa}}\right)\,,
 \nonumber\\*
 P_\perp&=&\sqrt{1-P_\parallel^2}\;
 \frac{2\kappa}{e^{\kappa}-e^{-\kappa}}\,,
\end{eqnarray}
where
\begin{equation}
\kappa \equiv \omega_0/\mu.
\end{equation}
For $\mu\to\infty$ and $\mu\to0$ the limits of $\omega_{\rm c}$ agree
with the results predicted in Eq.~(\ref{eq:boxlimits}). For
$\mu\to\infty$ we have $P_\perp=(1-P_\parallel^2)^{1/2}$,
representing the initial condition $P=1$. For $\mu\to0$ we find
$P_\perp=0$, corresponding to all polarization vectors either aligned
or anti-aligned with ${\bf B}$.

\begin{figure}[b]
\includegraphics[width=0.75\columnwidth]{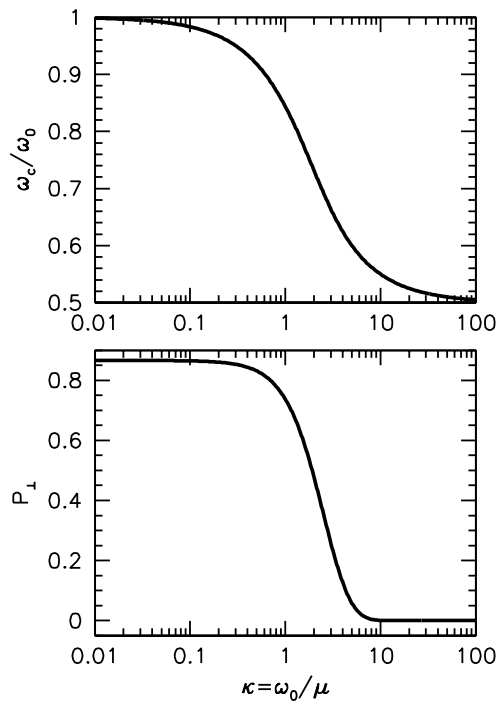}
\caption{Analytic results for $\omega_{\rm c}(\kappa)$ and
 $P_\perp(\kappa)$ according to Eq.~(\ref{eq:exact1}) with the initial
  condition $P_\parallel=0.5$.
\label{fig:analytic}}
\end{figure}

For our example with $P_\parallel=0.5$ we show $\omega_{\rm
  c}(\kappa)$ and $P_\perp(\kappa)$ in Fig.~\ref{fig:analytic}.
According to Eq.~(\ref{eq:exact1}), $\omega_{\rm c}$ decreases
monotonically with increasing $\kappa$ from $\omega_{\rm c} =
\omega_0$ to $\omega_{\rm c} = \omega_{\rm split} = 1 -
P_\parallel=0.5$. The transverse component $P_\perp$ decreases from
$(1 - P_\parallel^2)^{1/2}=\sqrt{0.75}=0.866$ down to $0$.
With Eq.~(\ref{eq:pz}) the analytic results provide exact adiabatic
solutions for the $\parallel$--component of the polarization-vector
spectrum.  In the bottom panel of Fig.~\ref{fig:spectrumevolution} we
show the evolution for 51~equally spaced modes in analogy to the
numerical result shown in the upper panel. The agreement for not very
small $\mu$ is striking, thus confirming the correctness of our
picture based on the adiabatic evolution in the co-rotating frame. The
curves obtained in the adiabatic approximation and by numerical
solution of the original evolution equations do not agree for the
modes close to the split ($\omega \approx \omega_{\rm split}$) at low
neutrino densities ($\mu < \omega_0$) where the evolution becomes
non-adiabatic.

Figure~\ref{fig:spectrumevolution} shows that the trajectories of
different modes literally split in the course of evolution, giving
us another (dynamical) motivation for the term ``spectral split'' to
describe the phenomenon.

It is remarkable that the split is essentially complete only for
$\omega_0/\mu\agt7$ or $\mu\alt0.15\,\omega_0$. In other words, the
overall motion remains collective in the sense of all polarization
vectors precessing in a common plane for a surprisingly small value
of $\mu$, even for imperfect adiabaticity as in the upper panel of
Fig.~\ref{fig:spectrumevolution}.

Actually for any value of $\mu$, no matter how small, our adiabatic
case is an exact solution of the EOMs where all polarization vectors
precess in a common plane. Of course, for very small $\mu$ their
angles relative to the positive or negative ${\bf B}$ direction are
exponentially small, but conceptually this is still a collective
motion. In the perfectly adiabatic case, collectivity is never lost.
Instead, the polarization vectors orient themselves in the positive
or negative ${\bf B}$--direction.

If $\mu$ were to switch off suddenly, the transverse component of the
overall ${\bf P}$ disappears by kinematical decoherence because all
${\bf P}_\omega$ precess with different frequencies. In the perfectly
adiabatic case, the transverse component disappears because all
individual ${\bf P}_\omega$ (anti)align themselves with ${\bf B}$
while remaining in a common plane. There is never any kinematical
decoherence caused by differences of precession frequencies between
different modes. The persistence of collectivity for arbitrarily small
$\mu$ is the most surprising aspect of this system.

\subsection{Spectral cross-over}                 \label{sec:crossover}

At the neutrino sphere in a SN it is expected that the $\nu_e$ flux
dominates for low energies and the $\nu_x$ flux for high energies.
This would be the case, for example, if both species are emitted with
equal luminosities but different average energies. In this case the
initial ${\bf P}_\omega$ are either aligned or anti-aligned with the
flavor direction. In the previous discussion we must then be careful
about absolute signs of the various projections of the polarization
vectors.

An important difference occurs in Eq.~(\ref{eq:hialigned}) because
${\bf P}_\omega$ can be anti-aligned with $\hat{\bf H}_\omega$ so
that $P_\omega$ should be taken negative for those modes that
initially contain the ``wrong'' flavor. The final state of these
modes will be opposite from where they would have ended if they had
been occupied with the ``correct'' flavor.

\begin{figure}[ht]
\includegraphics[width=1.0\columnwidth]{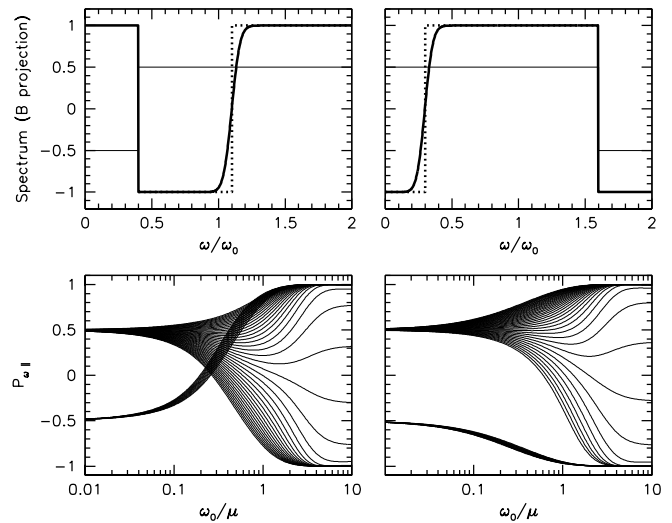}
\caption{Examples for crossed-over neutrino spectra. {\it Top:} Thin
  solid: initial. Thick dotted: adiabatic final. Thick solid:
  numerical final with $\tau=(0.03\,\omega_0)^{-1}$ in
  Eq.~(\ref{eq:muexp}). {\it Bottom:} Numerical evolution of
  51~equally spaced modes.\label{fig:cross1}}
\end{figure}

The final spectrum is again determined by flavor lepton-number
conservation. Imagine that the low-$\omega$ modes are initially in
the other flavor, corresponding to a fraction $\xi$ of all modes, and
assume a box spectrum (thin line in the top left panel of
Fig.~\ref{fig:cross1}). These modes must now end up at
$P_{\omega\parallel}=+1$, affecting the net flavor lepton number. The
modes above $\omega_{\rm split}$ also must end up at $+1$, whereas
intermediate ones end up at $-1$. It is now easy to work out that in
the end ($\mu\to0$)
\begin{equation}\label{omegacr}
 \omega_{\rm split} = \omega_0
 \left[1 - \cos \beta + 2\xi\,(1 +\cos\beta) \right],
\end{equation}
where $\beta$ is the initial angle between ${\bf B}$ and ${\bf P}$.
For $\xi = 0.2$ and $\cos\beta = 0.5$ we thus find $\omega_{\rm split}
= 1.1\,\omega_0$. In this way we predict for the example of
Fig.~\ref{fig:cross1} (left column) the final spectrum indicated by a
thick dotted line. Numerically we find the thick solid line.

The split at $\omega_{\rm split}$ is somewhat washed out because the
evolution is not fully adiabatic.  On the other hand, the transition
at the initial spectral cross over remains sharp because the end
state of those modes depends only on the initial sign of $P_\omega$,
but otherwise we are in an adiabatic regime of the spectrum.

In the bottom panel we show the familiar pattern of the
$P_{\omega\parallel}$ evolution with the novel feature that the modes
that were initially in the other flavor cross over to
$P_{\omega\parallel}=+1$.

In the right-hand panels of Fig.~\ref{fig:cross1} we show an
analogous example where the high-$\omega$ part of the initial
spectrum is crossed over. The explanation is analogous.

\section{Neutrinos and Antineutrinos}                 \label{sec:both}

\subsection{Solution for the double-box spectrum}

As a next case towards a more complete understanding we include
antineutrinos in our model, upholding the condition that all
polarization vectors ${\bf P}_\omega$, now with frequencies in the
range $-\infty<\omega<+\infty$, initially point in the same
direction. We assume that the fraction of antineutrinos is $\alpha<1$
of the neutrinos. We will always assume the inverted mass hierarchy so
that a large flavor transformation effect is caused by the instability
of the inverted flavor pendulum~\cite{Hannestad:2006nj, Duan:2007fw}.

As a generic example we extend the box spectrum of the previous
section to include antineutrinos. To be specific, we assume a
double-box spectrum of equal width, but different height for the
neutrinos and antineutrinos:
\begin{equation}
 P_{\omega}=\frac{1}{2\omega_0}\times
 \cases{\alpha&for $-2\omega_0\leq\omega\leq 0$,\cr
 1&for $0<\omega\leq 2\omega_0$,\cr
 0&otherwise.\cr}
\end{equation}
We will assume $\alpha=0.7$ as a standard case.

\begin{figure}[b]
\includegraphics[width=0.75\columnwidth]{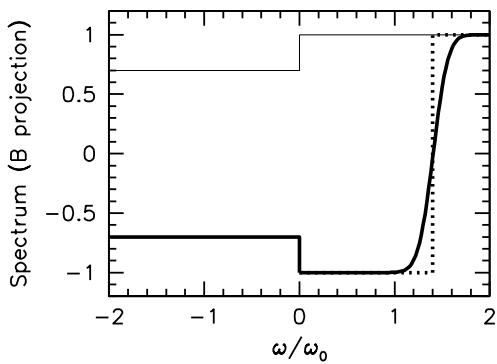}
\caption{Neutrino spectra for an initial box spectrum with 70\%
antineutrinos and $\sin2\theta_{\rm eff}^{\infty}=0.05$.
Negative frequencies correspond
to antineutrinos.  Thin line: initial. Thick dotted: final adiabatic.
Thick solid: numerical example with $\tau=(0.1\,\omega_0)^{-1}$ in
Eq.~(\ref{eq:muexp}).\label{fig:boxspectrum2}}
\end{figure}

The thin line in Fig.~\ref{fig:boxspectrum2} represents this initial
spectrum where antineutrinos are represented with negative
frequencies. Using the same arguments as in the neutrino-only case,
in the co-rotating frame the antineutrinos are modes with even more
negative frequencies than those neutrinos with $\omega<\omega_{\rm
split}$. Therefore, they should simply reverse their direction. This
is borne out by the numerical final spectra in
Fig.~\ref{fig:boxspectrum2} (thick solid line). The adiabatic limit
(thick dotted line) is again explained by the conservation of flavor
lepton number.

\begin{figure}
 \includegraphics[width=1.0\columnwidth]{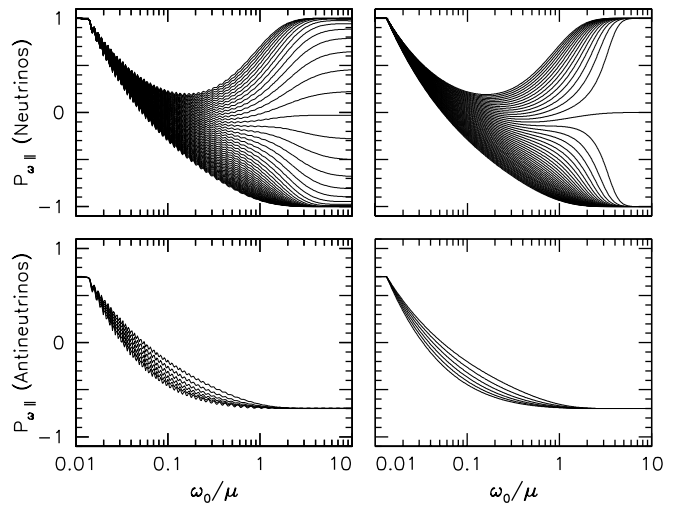}
\caption{$P_{\omega\parallel}(\mu)$ for individual modes for the case of
neutrinos plus antineutrinos. {\it Left:} Numerical solution from
the EOMs as in our previous paper~\cite{Raffelt:2007cb}
for $\sin2\theta_{\rm eff}^{\infty} =0.05$.
 {\it Right:} Adiabatic solution for $\sin2\theta_{\rm eff}^{\infty}=0$. In each
case neutrinos with 51 modes (top) and antineutrinos with 6 modes
(bottom).\label{fig:nunubarspectrum}}
\end{figure}

Next we consider the evolution of the individual modes.  In the upper
panels of Fig.~\ref{fig:nunubarspectrum} we show
$P_{\omega\parallel}(\mu)$ for neutrinos and in the lower panel for
antineutrinos. As expected, they evolve at first collectively and
both global polarization vectors tilt against the ${\bf B}$
direction, the familiar result of neutrino-neutrino refraction in the
inverted hierarchy case. Later the modes separate to produce the
final spectrum. In Fig.~\ref{fig:nunubarspectrum} we juxtapose the
direct numerical solution of the EOMs, assuming $\sin2\theta_{\rm eff}^{\infty}=0.05$
and a not very adiabatic $\tau=(0.1\,\omega_0)^{-1}$, with the fully
adiabatic solution for $\sin2\theta_{\rm eff}^{\infty}=0$ (right panels).

We could have illustrated this case by using the same length for the
${\bf P}_\omega$ for both neutrinos and antineutrinos, but occupying
only  70\% of the antineutrino modes. In this alternative double-box
case we could have shown the evolution of all modes in a single plot
because the antineutrino modes seamlessly join the neutrino modes.

A new feature of the left panels of Fig.~\ref{fig:nunubarspectrum} is
the ``wiggles'' in the curves that stem from the nutation of the
global flavor pendulum. We will return to this issue later but for
now only note that the wiggles disappear in the numerical solution if
we take the scale parameter $\tau$ very large,
much larger than is realistic in the SN
context. In this case the split would be extremely sharp. In order to
show a not fully adiabatic solution we have here used parameters that
do not have too many nutation periods (relatively small $\tau$) and
where the nutation depth is not too large (mixing angle not too
small) or else the plot would be too cluttered. In any case, there
are no nutations in an ``extremely adiabatic'' situation which for
the moment we focus on.

\begin{figure}[ht]
\includegraphics[width=0.4\textwidth]{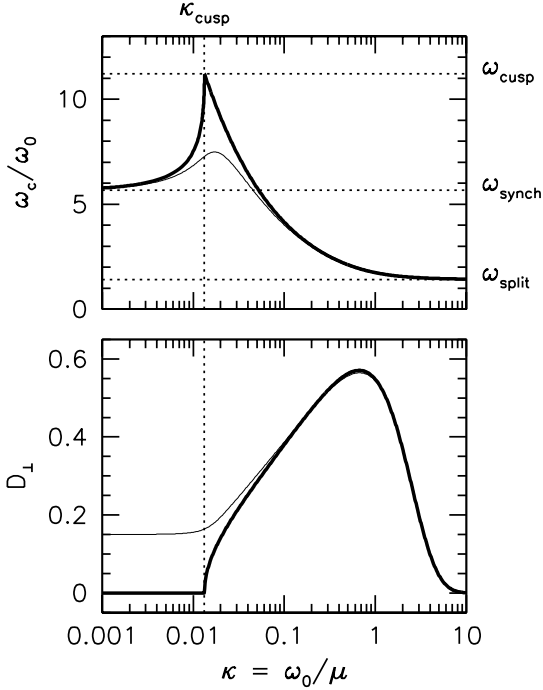}
\caption{Numerical solution of Eqs.~(\ref{eq:master2})
  and~(\ref{eq:master1}) for the double-box spectrum  of neutrinos plus
  antineutrinos with $\alpha=0.7$ and $\sin2\theta_{\rm eff}^{\infty}=0$. The vertical
  dotted line indicates $\kappa_{\rm cusp}=0.013340$ below which the
  system is perfectly synchronized. The horizontal dotted lines
  indicate $\omega_{\rm synch}=\omega_{\rm c}^\infty=(17/3)\,\omega_0$
  and $\omega_{\rm split}=\omega_{\rm c}^0=(14/10)\,\omega_0$. The
  thin solid line is the same case with
  $\sin2\theta_{\rm eff}^{\infty}=0.5$. \label{fig:numericalsolution}}
\end{figure}

The adiabatic solution for the individual modes shown in the
right-hand panels of Fig.~\ref{fig:nunubarspectrum} were found from
solving our sum rules Eqs.~(\ref{eq:master2}) and~(\ref{eq:master1})
numerically. The curves $\omega_{\rm c}(\mu)$ and $D_\perp(\mu)$ for a
vanishing mixing angle are shown as thick solid lines in
Fig.~\ref{fig:numericalsolution}.  It shows interesting new features
compared to the neutrino-only case. For $\mu\to\infty$ we have
$D_\perp=0$ due to the assumption of a vanishing mixing angle. The
precession frequency corresponds to the expected synchronization
frequency. However, for decreasing $\mu$ the precession frequency
increases up to an apparent cusp, and then decreases eventually down
to the expected split frequency. At the cusp, $D_\perp$ begins to
deviate from zero whereas $D_\perp=0$ for $0<\mu^{-1}<\mu_{\rm
  cusp}^{-1}$. In other words, the cusp marks the neutrino-neutrino
interaction strength where the polarization vectors begin to tilt
away from the ${\bf B}$ direction and begin to spread in the
zenith-angle direction.

For comparison we show, as a thin line, the same result for a large
initial mixing angle, $\sin2\theta_{\rm eff}^{\infty}=0.5$, implying
that initially $D_\perp=(1-\alpha)\,\sin2\theta_{\rm
eff}^{\infty}=0.15$.  The curve is qualitatively similar but does not
show any cusps or kinks. Notice that for $\omega_{\rm c}>2 \omega_0$
the co-rotation frequency is above the maximal frequency of the
spectrum for $\kappa < 1$. Therefore, a true split of the modes (in
contrast to a zenith-angle spread) only starts when $\omega_{\rm c} =
2 \omega_0$, that is at $\kappa = 1$. Apparently the dependences in
Fig.~\ref{fig:numericalsolution} differ from that in
Fig.~\ref{fig:analytic} for pure neutrino case.

\subsection{Properties of the pure precession mode}

Before turning to more physical interpretations of these features,
notably the cusp, we first investigate some mathematical properties
of these solutions. For the double-box spectrum, the sum rules can be
evaluated analytically. Eq.~(\ref{eq:master1}) gives
\begin{equation}
 \frac{\left( y_{+} + \sqrt{y_{+}^2 + D_{\perp}^2} \right)
       \left( y_{-} + \sqrt{y_{-}^2 + D_{\perp}^2} \right)^\alpha }{
       \left( y_{0} + \sqrt{y_{0}^2 + D_{\perp}^2} \right)^{1 + \alpha}
}
= e^{2\kappa},
\label{boxnubarnu1}
\end{equation}
where $y_{\pm} = y_0 \pm 2 \kappa$ and $y_0 \equiv D_\parallel -
(\omega_{\rm c}/\omega_0)\,\kappa$. The integration of
Eq.~(\ref{eq:master2}) leads to
\begin{equation}
\sqrt{y_{+}^2 + D_{\perp}^2}
+ \alpha \sqrt{y_{-}^2 + D_{\perp}^2}
- (1 + \alpha) \sqrt{y_{0}^2 + D_{\perp}^2}
= 2\kappa  D_\parallel\,.
\label{boxnubarnu2}
\end{equation}
Practically it is not possible to extract $D_\perp$ and $\omega_{\rm c}$
as functions of $\mu$
in closed form for generic values of parameters.

In a SN, the most interesting case is $\sin2\theta_{\rm eff}^{\infty} \approx 0$,
where
$D_\parallel=1-\alpha$.  This extreme case entails a number of
simplifications, but also a number of subtleties. It is easy to find
$\omega_{\rm c}(\kappa)$ for special values of $\kappa$.

\subsubsection{Synchronization frequency}

From Eq.~(\ref{eq:wsynch}) we easily derive the initial common
oscillation frequency, corresponding to the well-known synchronization
frequency,
\begin{equation}
 \omega_{\rm c}^\infty=\omega_{\rm synch}=
 \omega_0\,\frac{1+\alpha}{1-\alpha}\,.
\end{equation}
For $\alpha=\frac{7}{10}$ this is $\omega_{\rm synch}=
\frac{17}{3}\,\omega_0 = 5.667\,\omega_0 > 2\omega_0$. Therefore,
initially $\omega_{\rm c}$ is ``outside of the spectral box,'' i.e., all
neutrino and antineutrino modes have negative frequencies in the
co-rotating frame.

\subsubsection{Split frequency}

Later the modes split and the final split frequency at $\mu=0$ is
found from $\nu_e$ number conservation to be
\begin{equation}\label{critfr}
 \omega_{\rm split} =\omega_0\,(1 - D_z + \alpha)
 = \omega_0\,2\alpha\,,
\end{equation}
where we have used that for a very small mixing angle
$D_\parallel=1-\alpha$. With $\alpha= 0.7$ this is $\omega_{\rm split}
= 1.4 \,\omega_0$ in agreement with
Fig.~\ref{fig:boxspectrum2}. Independently of the choice of $\alpha$,
the final $\omega_{\rm split}$ always lies ``within the box'' because
$\omega_0\,2\alpha<2\omega_0$ so that a split always occurs.
In the end $D_\perp=0$ due to the (anti)alignment of all polarization
vectors with ${\bf B}$.

\subsubsection{Cusp}

A final special quantity is the strength of the neutrino-neutrino
interaction where the system exits the synchronization regime and
enters the ``bipolar regime.'' This transition is infinitely abrupt
when $\sin2\theta_{\rm eff}^{\infty}=0$ and corresponds to the cusp in
$\omega_{\rm  c}(\kappa)$ shown in Fig.~\ref{fig:numericalsolution}.

For $0<\kappa<\kappa_{\rm cusp}$ we note that $D_\perp=0$ solves
Eq.~(\ref{boxnubarnu2}) identically. Equation~(\ref{boxnubarnu1}) can
be written in the form
\begin{equation}\label{eq:ua}
(1-u)^\alpha(1+u)=e^{2\kappa}\,,
\end{equation}
where
\begin{equation}\label{eq:uk}
u=\frac{2\kappa}{D_\parallel-\kappa\,\omega_{\rm c}/\omega_0}\,.
\end{equation}
At $\kappa=\kappa_{\rm cusp}$ the function $\omega_{\rm
  c}(\kappa)$ appears to have a vertical tangent.  If this is true,
$u(\kappa)$ also should have a vertical tangent or $\kappa(u)$ should
have a horizontal one.  Equation~(\ref{eq:ua}) is trivially inverted
to provide an explicit expression for $\kappa(u)$. Using the condition
$d\kappa/du=0$ we find $u_{\rm cusp}$. Sticking this back into
$\kappa(u)$ we obtain
\begin{eqnarray}
 \kappa_{\rm cusp}&=&\frac{1}{2}
 \left[\log\left(\frac{2}{1+\alpha}\right)+
 \alpha\log\left(\frac{2\alpha}{1+\alpha}\right)
 \right]\,
 \nonumber\\*
 &=&0.01330\quad\hbox{(for $\alpha=0.7$).}
\end{eqnarray}
The co-rotation frequency at the cusp
is found to be
\begin{eqnarray}
 \frac{\omega_{\rm cusp}}{\omega_0}&=&
 \frac{1+\alpha}{\kappa_{\rm cusp}}-2\,\frac{1+\alpha}{1-\alpha}
 \nonumber\\*
 &=&11.2148\quad\hbox{(for $\alpha=0.7$)}.
\end{eqnarray}
The values of $\kappa_{\rm cusp}$ and $\omega_{\rm cusp}$ are in
perfect agreement with the numerical result shown in
Fig.~\ref{fig:nunubarspectrum}, thus confirming  that in the cusp
$\omega_{\rm c}$ has a vertical tangent.

\begin{figure}[b]
\includegraphics[width=0.9\columnwidth]{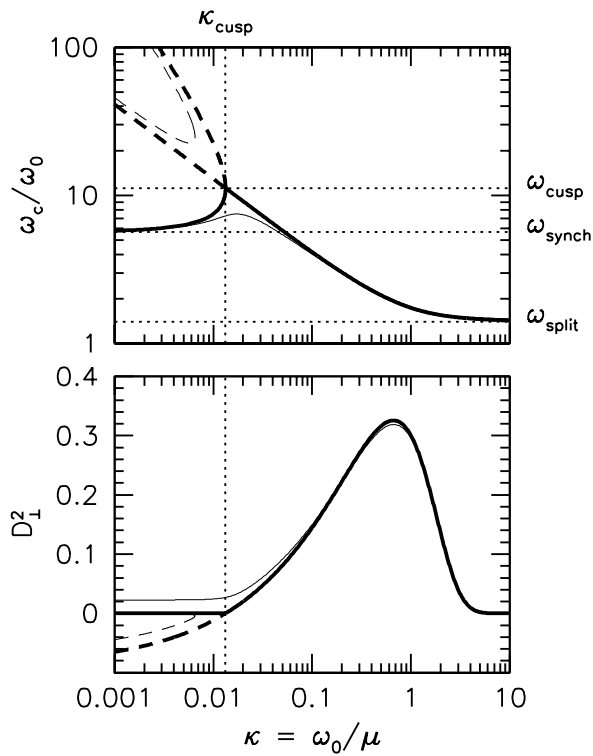}
\caption{Numerical solution of the sum rules Eqs.~(\ref{eq:master2})
  and~(\ref{eq:master1}) as in Fig.~\ref{fig:numericalsolution}, now
  including the unphysical solutions as dashed lines.
  \label{fig:numericalsolution2}}
\end{figure}

We observe that Eqs.~(\ref{eq:ua}) and~(\ref{eq:uk}) allow us to find
the curve $\omega_{\rm c}(\kappa)$ analytically in parametric form,
i.e., we can extract a pair of functions $\omega_{\rm c}(u)$ and
$\kappa(u)$. One easily shows that the function $\kappa(\omega_{\rm
c})$ does not end at the cusp, but rather is well defined for
$\omega_{\rm synch}\leq\omega_{\rm c}<\infty$. Turning this around,
the function $\omega_{\rm c}(\kappa)$ is double valued for
$0<\kappa<\kappa_{\rm cusp}$. We show the full solution in the upper
panel of Fig.~\ref{fig:numericalsolution2} where the unphysical part
is continued as a dashed line.

In addition, we find numerically that the high-$\kappa$ branch also
has a smooth continuation below the cusp. This branch is also
unphysical, corresponding to solutions with $D_\perp^2<0$ shown as a
dashed line in the lower panel of Fig.~\ref{fig:numericalsolution2}.
The cusp actually corresponds to a cross-over of two solutions,
although the physical solution indeed has a cusp.

The unphysical solutions persist for nonvanishing mixing angles. We
show the complete numerical solutions for $\sin2\theta_{\rm
  eff}^{\infty}=0.5$ as thin lines in
Fig.~\ref{fig:numericalsolution2}. The physical ($D_\perp^2>0$) and
unphysical ($D_\perp^2<0$) solutions are now fully separated: there
are no kinks or cusps as far as one can tell numerically.

For $\sin2\theta_{\rm eff}^{\infty}=0$ the physical solution has a
cusp, $D_\perp=0$ is exact for the entire range $0<\kappa<\kappa_{\rm
  cusp}$, yet $\omega_{\rm c}$ varies in this range as a function of
$\kappa$.  These surprising features are more easily understood in a
simpler case where one uses only two polarization vectors, one for
neutrinos and one for antineutrinos, i.e., for a monochromatic
neutrino spectrum. We will consider this example in
Sec.~\ref{sec:twovectors} and return there to a discussion of the
$\omega_{\rm c}$ behavior.

\section{Two Polarization Vectors}              \label{sec:twovectors}

As a last explicit example we study a system consisting of only two
polarization vectors ${\bf P}_1$ for neutrinos and ${\bf P}_2$ for
antineutrinos with $P_1=|{\bf P}_1|=1$ and  $P_2=|{\bf
P}_2|=\alpha<1$. This example allows us to make a connection to the
previous literature. Moreover, this system is a useful toy example to
understand some of the peculiarities of the general
neutrino-antineutrino system, notably the cusp feature in the common
precession frequency $\omega_{\rm c}(\mu)$. Finally, realistic
spectra do not extend to infinite energies and therefore do not have a
continuity in the region around $\omega =0$. In this respect the two
line-spectrum represent reality to a certain extend.

\subsection{Pure precession mode}

According to Eq.~(\ref{eq:eom1}) the equations of motion can be
written in the form
\begin{eqnarray}
 \dot{\bf P}_1&=&
 +\omega_0{\bf B}\times{\bf P}_1+\mu{\bf P}_1\times{\bf P}_2\,,
 \nonumber\\
 \dot{\bf P}_2&=&
 -\omega_0{\bf B}\times{\bf P}_2+\mu{\bf P}_1\times{\bf P}_2\,.
\label{twoveceq}
\end{eqnarray}
If ${\bf P}_1$ and ${\bf P}_2$ were free, both of them would precess
with frequency $\omega_0$, but in opposite directions.

We first investigate the ``pure precession mode''~\cite{Duan:2007mv}
of this system, at first without reference to adiabaticity, where
both polarization vectors precess in a single plane around ${\bf B}$.
If the vectors ${\bf B}$, ${\bf P}_1$ and ${\bf P}_2$ are indeed in a
single plane, then the velocity vectors being proportional to ${\bf
B}\times{\bf P}_1$, ${\bf B}\times{\bf P}_2$ and ${\bf P}_1\times{\bf
P}_2$ are all collinear. Let $\vartheta_1$ and $\vartheta_2$ be the
angles between ${\bf P}_1$ and ${\bf P}_2$ and ${\bf B}$,
respectively. If ${\bf P}_1$ and ${\bf P}_2$ are in the same plane,
then both evolution equations can be  reduced to the form $\dot{\bf
P}_i = \omega_{\rm c}{\bf B}\times{\bf P}_i$ ($i = 1$, 2) with
\begin{eqnarray}\label{eq:wcfuller}
 \omega_{\rm c}&=&+\omega_0+\mu\,\alpha\,
 \frac{\sin(\vartheta_2-\vartheta_1)}{\sin\vartheta_1}
 \nonumber\\*
 &=&+\omega_0+\mu\,\alpha\,\left(\frac{c_1s_2}{s_1}-c_2\right)\,,
 \nonumber\\*
 \omega_{\rm c}&=&-\omega_0+\mu\,
 \frac{\sin(\vartheta_2-\vartheta_1)}{\sin\vartheta_2}
 \nonumber\\*
 &=&-\omega_0+\mu\,\left(c_1-\frac{c_2s_1}{s_2}\right)\,,
\end{eqnarray}
where we have used the notation $c_1 \equiv \cos\vartheta_1$ and so
forth.  This is equivalent to Eq.~(65) of Duan et
al.~\cite{Duan:2007mv}.

Eliminating  $\mu\sin(\vartheta_2-\vartheta_1)$ from the first form
of these equations we find
\begin{equation}\label{eq:wc}
 \frac{\omega_{\rm c}}{\omega_0}=
 \frac{s_1+\alpha s_2}
 {s_1-\alpha s_2}\,.
\end{equation}
In the alignment case where $\vartheta_1=\vartheta_2$ we immediately
reproduce
\begin{equation}\label{eq:wsynch2}
 \frac{\omega_{\rm synch}}{\omega_0}=
 \frac{1+\alpha}{1-\alpha}
\end{equation}
for the synchronization frequency.

It is also trivial to eliminate $\omega_{\rm c}$ from the two
equations in Eq.~(\ref{eq:wcfuller}) and one finds
\begin{equation}\label{eq:kappa}
 \kappa\equiv\frac{\omega_0}{\mu}
 =\frac{s_1-\alpha s_2}{2}\,
 \left(\frac{c_1}{s_1}-
 \frac{c_2}{s_2}\right)\,.
\end{equation}
In the alignment case, $\vartheta_1=\vartheta_2$, the pure precession
mode requires $\kappa=0$ (or $\mu\to\infty$).

\subsection{Alignment with Hamiltonians}

It is surprising that for any two angles $\vartheta_1$ and
$\vartheta_2$ one finds solutions $\omega_{\rm c}$ and $\mu$ that
permit a pure precession. Of course, suitable initial conditions are
required to actually put the system into this mode. Demanding $\mu$
to be positive (as it would for neutrino-neutrino interactions)
implies some restrictions on the allowed angular ranges, but
otherwise all combinations of angles are possible.

As a check of self-consistency we note that in the pure precession
mode, ${\bf P}_{1,2}$ do  not move within the co-rotating frame.
Therefore, in this frame their Hamiltonians must be collinear with
${\bf P}_{1,2}$. We can write the EOMs in general as
\begin{eqnarray}
 \dot{\bf P}_1&=&{\bf H}_1\times{\bf P}_1\,,
 \nonumber\\
 \dot{\bf P}_2&=&{\bf H}_2\times{\bf P}_2\,.
\end{eqnarray}
The Hamiltonians in the co-rotating frame are
\begin{eqnarray}\label{eq:hamiltonians}
 {\bf H}_1&=&(+\omega_0-\omega_{\rm c}){\bf B}
 +\mu\left({\bf P}_1-{\bf P}_2\right)\,,
 \nonumber\\
 {\bf H}_2&=&(-\omega_0-\omega_{\rm c}){\bf B}
 +\mu\left({\bf P}_1-{\bf P}_2\right)\,.
\label{eq-for-hi}
\end{eqnarray}
They have angles relative to the ${\bf B}$--direction of
\begin{eqnarray}
 \cos\theta_1&=&
 \frac{\omega_0-\omega_{\rm c}+\mu c_1-\mu\alpha c_2}
 {\sqrt{(\omega_0-\omega_{\rm c}+\mu c_1-\mu\alpha c_2)^2
 +\mu^2(s_1^2-\alpha s_2)^2}},
 \nonumber\\*
 \cos\theta_2&=&
 \frac{-\omega_0-\omega_{\rm c} + \mu c_1-\mu\alpha c_2}
 {\sqrt{(-\omega_0-\omega_{\rm c} + \mu c_1-\mu\alpha c_2)^2
 +\mu^2(s_1^2-\alpha s_2)^2}}.
 \nonumber\\*
\end{eqnarray}
If we insert on the r.h.s.\ of each of these equations the expression
for $\omega_{\rm c}$ from the corresponding lines in
Eq.~(\ref{eq:wcfuller}) we find that indeed
$\cos\theta_{1,2}=\cos\vartheta_{1,2}$. Therefore, everything is
self-consistent: The pure precession mode is a form of motion of both
vectors where each is collinear with its Hamiltonian.

\subsection{Lepton-number conservation}

With some mild restrictions as noted above, the pure precession mode
is possible for almost any combination of $\vartheta_1$ and
$\vartheta_2$. In view of our application it is useful, however, to
group this two-dimensional set of solutions into families of
solutions that have a fixed ``lepton number,'' meaning solutions for
which
\begin{equation}
D_\parallel=\cos\vartheta_1-\alpha\cos\vartheta_2
\end{equation}
is a common fixed number. Each of these families includes the
alignment case $\vartheta_1=\vartheta_2\equiv\vartheta_0$ where
$D_\parallel=(1-\alpha)\,\cos\vartheta_0$. Therefore, we can write
the condition of lepton-number conservation in the form
\begin{equation}\label{eq:leptonconservation}
(1-\alpha)\cos\vartheta_0=\cos\vartheta_1-\alpha\cos\vartheta_2\,.
\end{equation}
In other words, once $\alpha$ has been chosen, we can classify the
solutions by the parameter $\vartheta_0$ or equivalently by
$D_\parallel$. All solutions with the same $\vartheta_0$ form a
family with the same lepton number and thus are adiabatically
connected to the aligned case where
$\vartheta_1=\vartheta_2=\vartheta_0$.

\subsection{Explicit solution}

We are now in a position to find explicit solutions of the EOMs in
the adiabatic limit where the two polarization vectors move in the
pure precession mode. To this end we specify a value for $\alpha$ and
$\vartheta_0$, an angle that has the interpretation of twice the
effective mixing angle. We can then use one of the tilt angles
$\vartheta_1$ or $\vartheta_2$ as a parameter to characterize
different solutions.

We know that the shorter anti-neutrino polarization vector ${\bf
P}_2$ for $\mu\to0$ will be anti-aligned with ${\bf B}$ whereas ${\bf
P}_2$ will remain at some finite angle fixed by lepton-number
conservation. Therefore, $\vartheta_2$ is a more useful independent
parameter. More specifically, we use
\begin{equation}
\eta \equiv \cos\vartheta_2
\end{equation}
as an independent variable that
varies between $\cos\vartheta_0\geq\eta\geq-1$. After eliminating
$\vartheta_1$ with the help of Eq.~(\ref{eq:leptonconservation}) it
is trivial to find from Eq.~(\ref{eq:wc}) explicitly
\begin{equation}\label{eq:wceta}
 \frac{\omega_{\rm c}}{\omega_0} = \frac{\sqrt{1-(c_0-\alpha
     c_0+\alpha\eta)^2} +\alpha\sqrt{1-\eta^2}} {\sqrt{1-(c_0-\alpha
     c_0+\alpha\eta)^2} -\alpha\sqrt{1-\eta^2}}\,,
\end{equation}
where $c_0\equiv\cos\vartheta_0=\cos2\theta_{\rm
  eff}^{\infty}$. Likewise, from Eq.~(\ref{eq:kappa}) we find
\begin{eqnarray}\label{eq:keta}
 \kappa&=&
 \frac{\sqrt{1-(c_0-\alpha c_0+\alpha\eta)^2}
 -\alpha\sqrt{1-\eta^2}}
 {2\sqrt{1-\eta^2} \sqrt{1-(c_0-\alpha c_0+\alpha\eta)^2}}
 \nonumber\\*
 &&{}\times
 \left[(c_0-\alpha c_0+\alpha\eta)\sqrt{1-\eta^2}\right.
 \nonumber\\*
 &&\kern5em
 \left.{}-\eta\sqrt{1-(c_0-\alpha c_0+\alpha\eta)^2}\right]\,.
\end{eqnarray}
While we cannot extract an explicit analytic result for $\omega_{\rm
c}(\kappa)$, we have an analytic form of this curve as $\omega_{\rm
c}(\eta)$ and $\kappa(\eta)$ in parametric form. Therefore, in
contrast to Ref.~\cite{Duan:2007mv} and in contrast to our
box-spectrum example we do not rely  on a numerical solution: All
properties of the solution can be understood analytically.

\begin{figure}[ht]
\includegraphics[width=0.9\columnwidth]{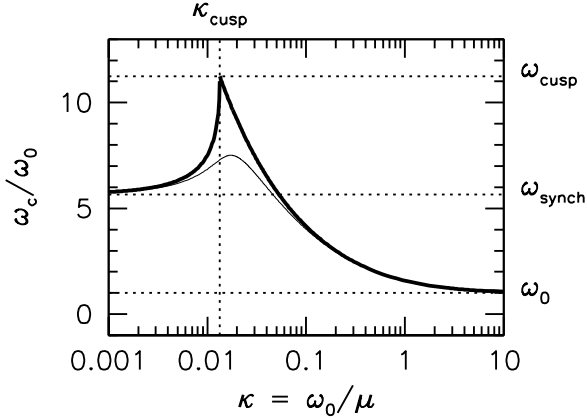}
\caption{Common precession frequency for our system of two
  polarization vectors with $\alpha=0.7$, in full analogy to the
  double-box spectrum case of Fig.~\ref{fig:numericalsolution} (upper
  panel).  The thick line is the limiting solution for
$\sin2\theta_{\rm eff}^{\infty}\to
  0$, equivalent to $\cos\vartheta_0\to 1$, the thin line is for
  $\sin2\theta_{\rm eff}^{\infty}=0.5$, corresponding to
  $\cos\vartheta_0=\sqrt{0.75}=0.866$ as in
  Fig.~\ref{fig:numericalsolution}.\label{fig:twopol}}
\end{figure}

In Fig.~\ref{fig:twopol} we show $\omega_{\rm c}(\kappa)$ as a thin
line for $c_0=\sqrt{0.75}=0.866$, corresponding to $\sin2\theta_{\rm
eff}^{\infty}=0.5$. This is equivalent to the thin line in the upper
panel of Fig.~\ref{fig:numericalsolution}, except that here we have
two polarization vectors whereas there we had a double-box spectrum.
The solutions are very similar. Using a smaller and smaller
$\sin2\theta$, corresponding to $c_0$ approaching 1, the solution
develops a sharper and sharper cusp as expected. We show the limiting
curve for $c_0=1$ or $\sin2\theta_{\rm eff}^{\infty}=0$ as a thick
solid line.

\subsection{Cusp}

The limiting case $\cos\vartheta_0\to 1$ involves a number of
subtleties. In the strictly adiabatic limit this solution corresponds
to both polarization vectors being strictly aligned with ${\bf B}$
or, in the neutrino language, to an exactly vanishing effective
mixing angle. Starting with this initial condition, reducing the
neutrino-neutrino interaction strength $\mu$ could never lead to a
deviation of the polarization vectors from this orientation just as
in the absence of mixing there can be no flavor conversions.

We first study this limiting case by using $c_0=1$ in
Eqs.~(\ref{eq:wceta}) and~(\ref{eq:keta}). The functions $\omega_{\rm
c}(\eta)$ and $\kappa(\eta)$ then give us only the large-$\kappa$
branch above the cusp. Specifically for $\eta\to 1$ we find the
familiar cusp values~\cite{Hannestad:2006nj, Duan:2007mv}
\begin{eqnarray}\label{eq:cusp2}
 \kappa_{\rm cusp}&=&\frac{(1-\sqrt\alpha)^2}{2}
 \nonumber\\*
 &=&0.01334\quad\hbox{for $\alpha=0.7$,}
 \nonumber\\*
 \frac{\omega_{\rm cusp}}{\omega_0}
 &=&\frac{1+\sqrt\alpha}{1-\sqrt\alpha}
 \nonumber\\*
 &=&11.2444\quad\hbox{for $\alpha=0.7$.}
\end{eqnarray}
Numerically these results are very similar to the double-box spectrum
case, but not identical.

Let us begin with two polarization vectors in the pure precession
mode with intermediate $\mu$ and let us assume the conserved
$D_\parallel$ is such that in the alignment case they have $c_0=1$.
If we now increase $\mu$ adiabatically, alignment with each other and
with ${\bf B}$ is reached for $\mu=\mu_{\rm cusp}$ so that
$\vartheta_1=\vartheta_2=0$. As we increase $\mu$ further until
infinity, they cannot get more aligned. This is different for all
cases with $c_0<1$ where the angles keep monotonically decreasing
with increasing $\mu$ and reach alignment
$\vartheta_1=\vartheta_2=\vartheta_0$ only for infinite $\mu$.

\subsection{Sleeping-top regime}

The range $\mu_{\rm cusp}<\mu<\infty$ in the limiting case has the
interpretation of the ``sleeping top regime''~\cite{Duan:2007mv}. In
the language of the gyroscopic flavor pendulum it corresponds to a
spinning top that spins so fast that energy and angular momentum
conservation prevent it from deviating from perfect vertical
alignment if it was prepared in this state, seemingly defying the
force of gravity. Therefore, we call this the sleeping-top range of
$\mu$ or $\kappa$.

We can study this regime using a somewhat different parametrization
that allows us to take the limit of a small angle $\vartheta_0$. To
this end we write $\vartheta_2=\xi\vartheta_0$ and thus express the
tilt angles as multiples of $\vartheta_0$. In Eqs.~(\ref{eq:wceta})
and~(\ref{eq:keta}) we thus substitute $c_0=\cos\vartheta_0$ and
$\eta=\cos\vartheta_2=\cos(\xi\vartheta_0)$. This is only a
re-parametrization: Once more we can find, for any chosen value of
$\vartheta_0$, the solution $\omega_{\rm c}(\kappa)$ in the
parametric form $\omega_{\rm c}(\xi)$ and $\kappa(\xi)$ with
$1\leq\xi\leq\pi/\vartheta_0$.

In the limit $\vartheta_0\to0$ we find for the co-moving frequency
\begin{equation}\label{eq:wcxi}
 \frac{\omega_{\rm c}}{\omega_0}
 =\frac{\sqrt{1+\alpha(\xi^2-1)}+\alpha\xi}
 {\sqrt{1+\alpha(\xi^2-1)}-\alpha\xi}\,.
\end{equation}
This leading term in a $\vartheta_0$ expansion does not depend on
$\vartheta_0$ and thus applies even for $\vartheta_0=0$. While in
this case $\vartheta_2=\xi\vartheta_0=0$, this result nevertheless
depends on $\xi$.

For $\xi=1$ we reproduce the expected synchronization value of
Eq.~(\ref{eq:wsynch2}) whereas for $\xi\to\infty$ we find
$\omega_{\rm cusp}$ of Eq.~(\ref{eq:cusp2}). In other words, in the
limit $\vartheta_0\to0$ our new parametrization covers the
sleeping-top branch, but not the region beyond the cusp.

One can take a similar representation for $\kappa(\xi)$ and use only
the leading term. This gives  the limiting values $\kappa(0)=0$ and
$\kappa(\infty)=\kappa_{\rm cusp}$ as expected.

Actually, Eq.~(\ref{eq:wcxi}) can be inverted to provide
$\xi(\omega_{\rm c})$ and the result can be inserted in the leading
term of $\kappa(\xi)$, providing after some simplifications
\begin{equation}
 \kappa=\frac{1}{\omega_{\rm c}/\omega_0+1}-
 \frac{\alpha}{\omega_{\rm c}/\omega_0-1}\,.
\end{equation}
This function is well defined in the entire range $\omega_{\rm
cusp}\leq\omega_{\rm c}<\infty$, i.e., once more we find both a
physical and an unphysical branch of the solution. We can also invert
this result and find for the physical branch
\begin{equation}
 \frac{\omega_{\rm c}}{\omega_0}=
 \frac{1-\alpha-\sqrt{\alpha^2-2\alpha(1+2\kappa)+(1-2\kappa)^2}}
 {2\kappa}\,,
\end{equation}
whereas the unphysical branch has a positive sign of the square root.
This function has real values only for $0\leq\kappa\leq\kappa_{\rm
cusp}$ and indeed approaches the familiar limits $\omega_{\rm synch}$
and $\omega_{\rm cusp}$ at the two ends of the sleeping-top interval.

\subsection{Sum rules}

The case of two polarization vectors studied in this section
corresponds to the spectrum
\begin{equation}
P_\omega=\delta(\omega-\omega_0)+\alpha\delta(\omega+\omega_0)\,.
\end{equation}
The adiabatic solutions must obey the sum rules discussed earlier.
For this two-line spectrum  Eqs.~(\ref{eq:master1})
and~(\ref{eq:master2aa}) give
\begin{eqnarray}
 1&=&\frac{1}{\sqrt{(\kappa-r\kappa+D_\parallel)^2+D_\perp^2}}
 \nonumber\\*
 &&\kern5em{}
 -\frac{\alpha}{\sqrt{(-\kappa-r\kappa+D_\parallel)^2+D_\perp^2}}\,,
\nonumber\\*
 r&=&\frac{1}{\sqrt{(\kappa-r\kappa+D_\parallel)^2+D_\perp^2}}
 \nonumber\\*
 &&\kern5em{}
 +\frac{\alpha}{\sqrt{(-\kappa-r\kappa+D_\parallel)^2+D_\perp^2}}\,,
\end{eqnarray}
where $r \equiv \omega_{\rm c}/\omega_0$. For any two angles
$\vartheta_1$ and $\vartheta_2$ we can calculate the corresponding
$D_\parallel$, $D_\perp$, $\omega_{\rm c}$ and $\kappa$ and show that
the sum rules are actually fulfilled.

In the general case we used the sum rules to find solutions
$\omega_{\rm c}(\kappa)$ and $D_\perp(\kappa)$ and then find
solutions for the individual polarization vectors which here would
mean to find $\vartheta_1(\kappa)$ and $\vartheta_2(\kappa)$. In the
two-polarization vector case we were able to proceed in the opposite
way and find, for a given $\vartheta_0$ (or a given $D_\parallel$)
the solutions $\vartheta_1$, $\omega_{\rm c}$ and $\kappa$ as a
function of $\vartheta_2$. This was possible because for two
polarization vectors lepton-number conservation uniquely fixes the
angle $\vartheta_1$ as a function of $\vartheta_2$.

\subsection{Spectral split?}

For two polarization vectors a spectral split cannot occur because it
would amount to the polarization vector ${\bf P}_1$ breaking apart.
The true final state consists of ${\bf P}_2$ indeed anti-aligning
with ${\bf B}$ and ${\bf P}_1$ retaining a non-zero transverse
component. The ${\bf P}_1$ final orientation is determined by
lepton-number conservation. In the end it precesses freely around
${\bf B}$ with frequency $\omega_0$. Indeed, we have found that in
the end $\omega_{\rm c}^0=\omega_0$.

At the same time we have noted that in the pure precession mode the
polarization vectors are aligned with their Hamiltonians which in the
end seem to be (anti)parallel to ${\bf B}$ because $\mu=0$. Notice,
however, that ${\bf H}_1$ in Eq.~(\ref{eq:hamiltonians}) involves
$(\omega_0-\omega_{\rm c}){\bf B}$ and if in the end $\omega_{\rm
c}\to\omega_0$ the final length of ${\bf H}_1$ approaches zero. In a
limiting sense it always maintains a direction tilted against ${\bf
B}$ such that ${\bf P}_1$ always remains aligned with ${\bf H}_1$.

Therefore, we here have a case where the perfectly adiabatic
evolution leads to a final configuration where one of the
polarization vectors is not (anti)aligned with ${\bf B}$.

For an arbitrary number of discrete polarization vectors an exact
spectral split will usually not be possible because typically one of
vectors would have to be broken apart to achieve a split. In this case
the final $\omega_{\rm c}$ will be the $\omega$ of the left-over
polarization vector that can neither align nor anti-align with ${\bf
B}$ if lepton number is to be conserved. Therefore, we expect that in
this situation the final adiabatic solution will be such that all
polarization vectors (anti)align with ${\bf B}$ except for this
remaining one that sits on the fence and for which in the end
$(\omega-\omega_{\rm c})\to 0$. In this way it remains aligned with
its Hamiltonian alright. We have seen such a case in the fully
adiabatic example in Fig.~\ref{fig:nunubarspectrum} where one mode
ended up at $P_{\omega\parallel}=0$.

\section{Adiabaticity violation}
\label{sec:transients}

\subsection{Sharpness of the spectral split}
\label{sec:sharpness}

In practice, $\mu(t)$ decreases with a finite speed. On the other
hand, the ${\bf P}_\omega$ with frequencies close to $\omega_{\rm
split}$ precess very slowly at late times when $\mu\to0$, so that
adiabaticity is broken for those modes as discussed earlier. As a
consequence, the spectral split is not infinitely sharp. There is a
transition region of non-zero width. A quantitative
estimate of the adiabaticity condition is given in our earlier
paper~\cite{Raffelt:2007cb}.

The width of the transition region, i.e., the sharpness of the
spectral split, is a common measure of the degree of adiabaticity
also for the ordinary MSW effect where the split occurs at
$\omega=0$.

\begin{figure}[ht]
\includegraphics[width=0.7\columnwidth]{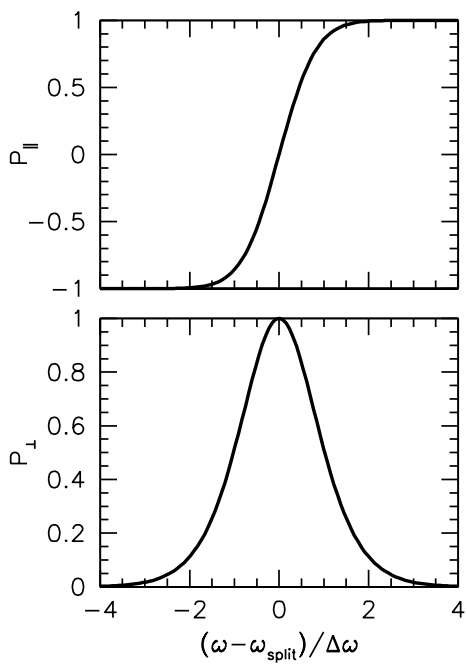}
\caption{Final spectra of $P_\parallel(\omega)$ (upper
panel) and $P_\perp(\omega)$ (lower panel) for a double-box spectrum
  of neutrinos plus antineutrinos as in Sec.~V. Here, $\Delta\omega$
  is the rms width of the distribution which depends on the time scale
  $\tau$ for the exponential $\mu(t)$ variation.
\label{fig:sharpness}}
\end{figure}

In order to quantify the meaning of ``sharpness'' we consider the
example of neutrinos plus antineutrinos with a double-box spectrum
with the average frequency $\omega_{0}$  as in Sec.~V and study the
final neutrino spectrum. From Fig.~7 it is evident that the
(anti)alignment of all polarization vectors with ${\bf B}$ is almost
perfect, even if adiabaticity is violated, except for the modes
around the split. In Fig.~\ref{fig:sharpness} we show the
final $P_\parallel(\omega)$ and $P_\perp(\omega)$ as a function of
$(\omega-\omega_{\rm split})/\Delta\omega$. Here, $\Delta\omega$ is
the variance or root-mean square width of the distribution
$P_\perp(\omega)$. Of course, since $P_\omega$ is conserved both
distributions contain the same information, but $P_\perp$ lends
itself more directly to a straightforward definition of the width of
the non-adiabatic range.

In this form, the shown functions appear to be universal for the
assumed double-box spectrum as far as we can tell from a numerical
exploration of parameters. In particular, these functions do not seem
to depend on the assumed asymmetry parameter $\alpha$, not on the
value of $\omega_{\rm split}$ and not on the mixing angle. They do not
depend on the degree of adiabaticity, i.e., on the time scale $\tau$
of the exponential $\mu(t)$ variation, except through
$\Delta\omega$. Varying $\omega_0\tau$ over two decades between $10$
and $10^3$ we find that
\begin{equation}
\Delta\omega  \approx \omega_0\, \frac{1}{(\omega_0\tau)^{3/4}}
\label{tau-deps}
\end{equation}
is an excellent representation with an accuracy better than a few
percent. (For $\omega_0\tau=100$ this implies
$\Delta\omega=0.032\,\omega_0$.) However, this relation is not exact:
The dependence of $\Delta\omega$ on $\tau$ does not seem to be an
exact power law. We give some explanations of these features in the
following subsection.

\subsection{MSW effect vs.\ collective transformations}

To shed more light on the dynamics of the split phenomenon we compare
here the adiabaticity violation with its counterpart in the usual MSW
case where the evolution is given by Eq.~(\ref{eq:eom1}) with $\mu =
0$ and where the EOMs for different frequencies $\omega$ decouple.

Let us consider first the MSW case for the normal mass hierarchy. If
the initial density is very large, much larger than the resonance
density, $\lambda \gg \lambda_{\rm res}(\omega)$, the initial
effective mixing is strongly suppressed. The adiabatic evolution to
small densities, $\lambda \to 0$, then leads to the well-known strong
conversion with a final survival probability $P = \sin^2
\theta\approx0$. The neutrino polarization vector initially coincides
with the Hamiltonian. Following the Hamiltonian completely flips the
flavor. With increasing energy $E$ (decreasing frequency $\omega$)
the degree of adiabaticity decreases and above a certain energy
$E_{\rm na}$ the effects of adiabaticity violation increase
exponentially. $P$ increases and approaches~1. In the antineutrino
channel for $E \rightarrow \infty$ ($\omega \rightarrow 0$) the
mixing is strongly suppressed and the survival probability $\bar{P}
\approx 1$.  So, there is a continuity in the $\omega$--variable
around $\omega \sim 0$, i.e., at the transition from neutrinos to
antineutrinos.

The split frequency can be estimated using the Landau-Zener
probability which describes well the survival probability for small
mixing angles~\cite{Haxton:1986dm, Parke:1986jy}: $P \approx P_{\rm
LZ} = e^{-A\omega}$. Here $A \equiv \pi \tau \sin^2 2\theta/(2\cos
2\theta)$, and $\tau = \lambda/\dot{\lambda}$ is the scale height of
the density change. From the condition $P_{\rm LZ} = 1/2$ we find
\begin{equation}
\omega_{\rm split} =
\frac{2\ln 2}{\pi \tau}\frac{\cos 2\theta}{\sin^2 2\theta}\,.
\label{msw-split}
\end{equation}
For the width of the split region one finds $\Delta \omega \sim
\omega_{\rm split}$ and $\Delta \omega /\omega_{\rm split}$ is a
universal quantity that does not depend on the parameters of the
problem. According to Eq.~(\ref{msw-split}) the sharpness of the
split is
\begin{equation}
\Delta \omega \propto \frac{1}{\tau}
\label{sharp-msw}
\end{equation}
which  differs from the dependence in Eq.~(\ref{tau-deps}) for the
collective transformation case.

We emphasize that here as well as in the collective transformation
case the split phenomenon requires a specific initial condition: a
small effective mixing angle.

For the MSW case in the normal mass hierarchy the split is in the
neutrino channel and the flavor flips for $\omega > \omega_{\rm
split}$. For the inverted hierarchy the split is in the antineutrino
channel and the modes with $\omega < \omega_{\rm split}$ flip.

With an increasing degree of adiabaticity (increasing $\tau$) the
split frequency shifts to smaller values, approaching $\omega_{\rm
  split} = 0$ which is realized in the ideally adiabatic
case. Conversely, with a decreasing degree of adiabaticity both
$\omega_{\rm split}$ and the width of the transition region
increase. This differs from the collective transformation case where
$\omega_{\rm split}$ does not shift and is determined solely by
lepton-number conservation. In other words, the changes of
$\omega_{\rm c}$ and $D_\perp$ adjust themselves self-consistently in
such a way that $\omega_{\rm split} = {\rm const}$. In the MSW case
flavor lepton number is not conserved.

Let us consider the origin of the differences between these two
cases. The Hamiltonian $H_\omega = {\bf H}_{\omega}\cdot{\bm\sigma}$,
where $\bm\sigma$ is the vector of Pauli matrices, has the diagonal
and off-diagonal elements $\pm H$ and $\bar{H}$, respectively,
\begin{equation}
2H = \mu D_\parallel - (\omega - \omega_{\rm c}), \qquad
2\bar{H} = \mu D_{\perp}\,.
\label{elements}
\end{equation}
The resonance condition, $H = 0$, can be written as
\begin{equation}
\omega = \mu_{\rm res} D_\parallel + \omega_{\rm c} (\mu_{\rm res}).
\end{equation}
The general form of the adiabaticity condition 
is $\dot{\theta}_{\rm eff} \ll \Delta H$, where $\theta_{\rm
eff}$ is the effective mixing angle and $\Delta H$ the difference of
eigenvalues. In the resonance region 
this condition takes on the form
\begin{equation}
\frac{\dot{H}}{4 \bar{H}^2} \ll 1\,.
\label{general}
\end{equation}
Using Eq.~(\ref{elements}) we find explicitly
\begin{equation}
\frac{D_\parallel +
\frac{d \omega_{\rm c}}{d\mu}}{2\tau_{\mu}\mu_{\rm res}
D_{\perp}(\mu_{\rm res})} \ll 1\,.
\label{adiab-self}
\end{equation}
The transition to the usual MSW case is given by
\begin{eqnarray}
\mu &\rightarrow& \lambda \, ,
\nonumber\\*
(\omega - \omega_{\rm c}) &\rightarrow& \omega \, ,
\nonumber\\*
D_\parallel &\rightarrow&L_\parallel = L \cos 2\theta \, ,
\nonumber\\*
D_\perp &\rightarrow& L_\perp = L \sin 2\theta .
\end{eqnarray}
A substantial difference is that in the MSW case both $L_\parallel$
and $L_\perp$ are constants and determined by the vacuum mixing angle.
Moreover, $\omega_{\rm c}$ is absent. The resonance condition in the
mass basis is: $\omega  = \lambda_{\rm res}\cos 2 \theta$. The
adiabaticity condition in the resonance becomes
\begin{equation}
\frac{1}{2\tau_\lambda \omega \sin^2 2\theta \cos 2\theta} \ll 1 \, . 
\label{adiab-msw}
\end{equation}

An essential difference of the collective case is that $D_\perp$ is a
dynamical variable and as such depends on $\mu$. Therefore the
effective mixing angle changes not only due to the shrinking of the
vector $\mu {\bf D}$ in analogy to $\lambda {\bf L}$ in the MSW case,
but also due to the rotation of ${\bf D}$. Consequently the width of the
resonance layer, where $2\theta_{\rm eff}$ changes from $45^{\circ}$
to $135^{\circ}$, and the adiabaticity condition essentially do not
depend on the initial mixing angle. This explains the universality of
the function that describes the spectral split.

The l.h.s.\ of the inequalities~(\ref{adiab-self})
and~(\ref{adiab-msw}) are adiabatic parameters which, up to numerical
coefficients, give the parameters in the exponents of the
corresponding Landau-Zener probabilities. They determine how the
sharpness of the spectral split depends on various parameters.
Apparently Eq.~(\ref{adiab-msw}) reproduces the result
Eq.~(\ref{sharp-msw}). In the collective case the dependence of
$\Delta \omega$ on $\tau_\mu$ is more complicated due to the presence
of $\omega_{\rm c}$ and $D_\perp$ which depend on $\mu$. These
functions are given in Fig.~\ref{fig:numericalsolution}.
Approximating these functions with power law functions in the
resonance region and inserting them into Eq.~(\ref{adiab-self}) can
provide an explicit expression for the shape of the washed-out
spectral split.

\subsection{Initial alignment of polarization vectors}

When we identified a sequence of static solutions with fixed lepton
number and variable $\mu$ as the adiabatic solution for a slow
$\mu(t)$ evolution, we assumed that initially $\mu\to\infty$. We
assumed that initially all neutrinos are prepared in weak interaction
eigenstates, corresponding to all ${\bf P}_\omega$ being aligned in a
direction that is tilted with the angle $2\theta_{\rm eff}^{\infty}$
relative to ${\bf B}$. This initial condition of perfect alignment
corresponds to a pure precession solution only if $\mu\to\infty$.

In a realistic system $\mu$ cannot be infinite. Therefore, if we
begin with an initial condition of perfect alignment combined with a
finite $\mu$, the system is initially not in a pure precession mode.
Even if the subsequent $\mu(t)$ evolution is infinitely slow, the
polarization vectors will not be perfectly aligned with the
Hamiltonians of the adiabatic solution.

In a linear system this situation would not spell a deviation from
adiabaticity. Usually adiabaticity is taken to mean that the
polarization vectors follow the Hamiltonians, but they need not be
aligned with them. Rather, the precession cones would retain a
non-zero but fixed opening angle.

Here the situation is different in that the ${\bf P}_\omega$ define
${\bf D}$ and thus affect the single-mode Hamiltonians ${\bf
H}_\omega$. Therefore, these Hamiltonians must now themselves show
fast motions when they are viewed from the co-rotating frame. Still,
if the initial $\mu$ is large (but finite), the deviation from a
static solution is small and one expects that nonlinear effects
appear only at second order in the small deviations of the
polarization vectors from their adiabatic solutions. One would expect
that the ${\bf P}_\omega$ precess around the nearly static ${\bf
H}_\omega$ where the opening angle of the precession cone no longer
vanishes. Notice that since ${\bf D}$ depends on many modes the fast
motion of the latter to some extend cancels in ${\bf D}$ and
therefore ${\bf H}_\omega$. This behavior is borne out in numerical
examples.

This numerical observation implies that the adiabatic solution is a
stable fixed point of the system, i.e., a small perturbation does not
lead to a run--away. Rather, the system always evolves close to the
adiabatic solution and oscillates around it. The stability of the
adiabatic solution is not automatically implied by its existence as a
self-consistent solution of the EOMs. Since we do not have developed
a formal criterion of stability, we do not know if there could be
examples that would not be stable. Therefore, it is a numerical
observation, not a proven fact, that the adiabatic solution is a good
approximation to the exact solution for typical cases of realistic
initial conditions.

\subsection{Nutations}

\begin{figure}[ht]
\includegraphics[width=1.0\columnwidth]{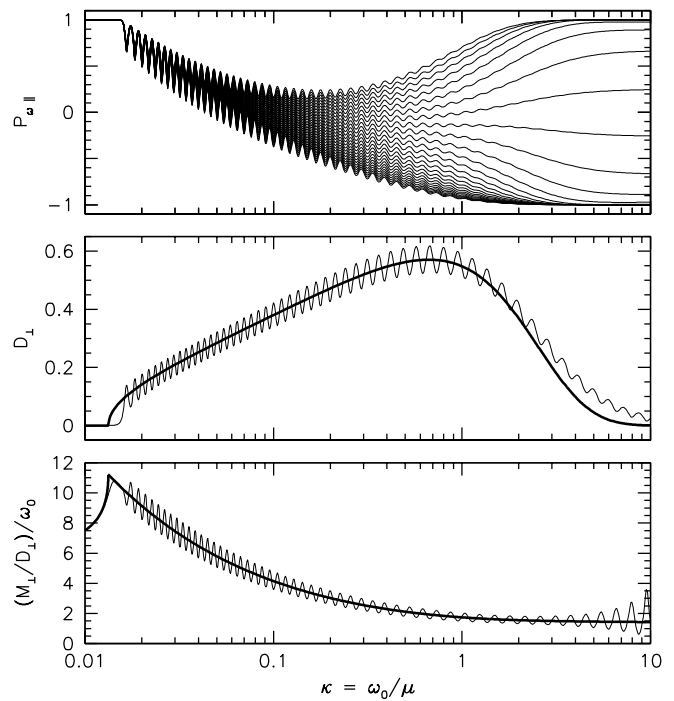}
\caption{Top: $P_{\omega\parallel}(\mu)$ for 25~individual modes for
the case of neutrinos plus antineutrinos. Same as upper left panel of
Fig.~\ref{fig:nunubarspectrum}, but now with a small mixing angle
$\sin2\theta_{\rm eff}^{\infty}=10^{-3}$. Middle: $D_\perp$ calculated from the
numerical solution compared with the adiabatic solution (thick line). Bottom:
$M_\perp/D_\perp$ in units of $\omega_0$ calculated from the
numerical solution compared with the adiabatic $\omega_{\rm c}$
(thick line).\label{fig:nunubarspectrum2}}
\end{figure}

A system consisting of neutrinos and antineutrinos is similar to the
neutrino-only case if one interprets the antineutrinos as
negative-frequency modes, but it has a number of peculiar properties.
One such a feature consists of the ``wiggles'' in the
$P_{\omega\parallel}$ evolution that we noted in the numerical
solution of the EOMs shown in Fig.~\ref{fig:nunubarspectrum}. The
amplitude of these modulations becomes deeper if the effective  mixing
angle is smaller. To illustrate this point we show in
Fig.~\ref{fig:nunubarspectrum2} (top panel) the same example as in the
upper-left panel of Fig.~\ref{fig:nunubarspectrum}, now with a smaller
mixing angle $\sin2\theta_{\rm eff}^{\infty}=10^{-3}$.

Our general treatment shows that in the limit of a very slow $\mu(t)$
variation and if initially $\mu\to\infty$ all polarization vectors
move in a co-rotating plane where the EOMs are solved by our
adiabatic solution without wiggles. They are, therefore, a
non-adiabatic feature and indeed disappear numerically if we use an
unrealistically slow $\mu(t)$ variation.

It has been shown that the EOMs of a system consisting of only two
polarization vectors, one for neutrinos and one for anti-neutrinos as
in Sec.~\ref{sec:twovectors}, are equivalent to a gyroscopic pendulum
in flavor space~\cite{Hannestad:2006nj, Duan:2007fw}. Its general
motion consists of a precession around ${\bf B}$ and a nutation that
causes the overall zenith-angle modulation. For very large
neutrino-neutrino interactions (very small $\kappa$), the pendulum is
in the ``sleeping top phase'' \cite{Duan:2007fw} where its motion is
a nearly pure precession. In the neutrino language this corresponds
to the synchronized regime where all polarization vectors are
strictly pinned to each other and precess around ${\bf B}$ as a
collective object. Once $\kappa$ becomes larger than the ``cusp
value,'' the gyroscope begins to wobble. In the neutrino language,
this is the point where the polarization vectors begin to spread in
the zenith-angle direction and are no longer strictly pinned to each
other.

The nutations are excited because the flavor pendulum initially
deviates only by the small angle $2\theta_{\rm eff}^{\infty}$ from
the inverted vertical direction. Starting the pendular motion
requires this small angle to grow under the influence of the external
force, introducing a time scale that depends logarithmically on
$2\theta_{\rm eff}^{\infty}$. If the $\mu(t)$ variation is fast
compared to this time scale, the pendulum cannot follow the adiabatic
solution. This explains that the depth of the nutation amplitude
becomes larger for a smaller $2\theta_{\rm eff}^{\infty}$. The case
of a vanishing $2\theta_{\rm eff}^{\infty}$ was the main example that
we studied in Sec.~\ref{sec:both}. While one can find a perfectly
self-consistent static solution for this case, it is clear that the
system cannot follow this solution: For a vanishing mixing angle the
system will not move at all, i.e., the delay leading to nutations is
infinite. Therefore, this example was an instructive limiting case,
but cannot be realized as an adiabatically connected sequence of
static solutions.

A remarkable feature of the nutations shown in
Fig.~\ref{fig:nunubarspectrum2} is that they are a collective effect
of all modes. We cannot interpret the nutations as a simple
precession of the individual ${\bf P}_\omega$ around the static
Hamiltonians. Rather, the nutations represent a motion of the entire
system around the adiabatic solution. To illustrate this point we
show in the middle panel of Fig.~\ref{fig:nunubarspectrum2} the
evolution of the numerical $D_\perp$ in comparison with the
corresponding adiabatic result. The true solution oscillates around
the adiabatic solution with the nutation period. Likewise, the
instantaneous precession frequency of the co-rotating plane, given by
the quantity $M_\perp/D_\perp$, oscillates around the adiabatic
result for $\omega_{\rm c}$ (bottom panel).

Once more we are numerically led to the conclusion that the adiabatic
solution represents a stable attractor of the nonlinear system. The
true solution is a collective oscillation mode around this solution
and thus plays the role of a perturbation. At the same time we stress
that the individual ${\bf P}_\omega$ evolve under the influence of
their instantaneous Hamiltonians ${\bf H}_\omega$ which are not
identical with the ones of the adiabatic solution. The instantaneous
Hamiltonians themselves nutate, although they nutate around the
static Hamiltonians. We recall, however, that the instantaneous
Hamiltonians always lie in a single plane, whereas this is not the
case for the polarization vectors. Moreover, the motion of the
instantaneous Hamiltonians is governed by the variation of the single
vector ${\bf D}$ so that they must move in lockstep with each other.

Another important feature of the nutations is that they are a
transient phenomenon that dies out toward the end of the evolution
when $\mu\to0$ as explained in the flavor pendulum
picture~\cite{Hannestad:2006nj, Duan:2007fw}. As a consequence, the
spectral split remains unaffected. Even though all polarization
vectors wildly nutate, they begin and end (anti)aligned with ${\bf
B}$ except for those modes near the split that anyway do not reach
the final (anti)alignment.  The modes sufficiently far from the split
begin and end aligned with the static Hamiltonians, whereas for
intermediate times they wildly nutate. The nutations have no apparent
impact on the final flavor spectrum of the system.

\section{Conclusion}                            \label{sec:conclusion}

We have elaborated our previous theory~\cite{Raffelt:2007cb} of
spectral splits that occur as a consequence of the adiabatic
evolution of a neutrino ensemble. We have clarified that the
spectral-split phenomenon in itself is not an entirely  new effect,
but a very useful way of looking even at the usual MSW effect if we
express its energy dependence in terms of the oscillation frequency
$\omega=\Delta m^2/2E$ and if we identify antineutrino modes with
negative frequencies. However, in the case of collective
transformations the effect has some distinctive features: the
position of the split is fixed by the flavor lepton number of the
system and does not depend on other parameters.

The main impact of collective neutrino-neutrino interactions is that
the evolution is adiabatic not in the laboratory frame, but rather in
a co-rotating frame that is defined by a plane in flavor space which
is spanned by the mass direction ${\bf B}$ and the vector ${\bf D}$
which is the polarization vector of the overall flavor lepton number
of the system. This frame defines a static solution of the EOMs that
consists, in the laboratory frame, of a pure precession of all
polarization vectors around ${\bf B}$. If the evolution in the
co-rotating frame is adiabatic, we will have a sharp spectral split
at the frequency $\omega=0$ in the co-rotating frame, corresponding
to a frequency $\omega_{\rm split}$ in the laboratory frame where
usually $\omega_{\rm split}\not=0$.

In order to take advantage of this picture we must find the
co-rotating frame explicitly. This is achieved in terms of two sum
rules ($\omega$--integrals over the spectrum $P_\omega$) that
guarantee the self--consistency of the static solution, i.e., of the
pure precession mode.

The main part of our paper is devoted to explicit analytical and
numerical solutions of these equations (sum rules) for generic
examples.  The adiabatic solution is determined completely by two
functions $\omega_{\rm c}(\mu)$ and $D_\perp (\mu)$. We study
properties of these functions for several different neutrino
spectra. For the neutrino-only box spectrum and non-zero initial
effective mixing angle both functions decrease monotonically with
$\mu$: $\omega_{\rm c}(\mu) \rightarrow \omega_{\rm split}$ and
$D_\perp (\mu) \rightarrow 0$.

In the presence of antineutrinos the behavior substantially changes:
For small initial mixing angles the dependence $\omega_{\rm c}(\mu)$
has a cusp at a certain density $\mu_{\rm cusp}$. The cusp corresponds
to the border between the sleeping-top region and the beginning of
bipolar transformations. Essentially the flavor evolution starts for
densities below the cusp. Above the cusp, $\omega_{\rm c}(\mu)$
increases starting from the synchronization frequency. Below the cusp
it decreases monotonically.  With decreasing $\mu$ we find that
$D_\perp(\mu)$ first increases, reaches a maximum at $\mu \sim
\omega_0$, a typical frequency of the spectrum, and then decreases.

This type of behavior appears to be universal and shows up for
completely different spectra. Very similar quantitative features
appear for the continuous double-box spectrum and for the
bi-chromatic spectrum. The latter case, by the gyroscopic
flavor-pendulum analogy, allows one to understand a number of subtle
features which also show up for continuous spectra.

The adiabatic solution is an approximate solution of the EOMs. There
are several features which appear in the exact solution that are not
described by the adiabatic solution. The violation of adiabaticity
always takes place at late stages of evolution  (at small $\mu$) for
modes with $\omega$ close to  $\omega_{\rm split}$. This leads to a
partial wash-out of the split so that the degree of adiabaticity
determines the sharpness of the split. Numerically it appears that
for a given spectrum the profile of the split is a universal function
of the density gradient and does not seem to depend on other
parameters. Measuring the width of the split region would allow one
to determine the neutrino density  gradient in the region $\mu \sim
\omega_0$.

Nutations related to bipolar transitions represent another deviation
from adiabaticity. The nutations occur around the adiabatic solution,
or, in other words, the adiabatic solution appears as an average over
nutations. Somewhat surprisingly, the nutations do not visibly
influence the final result.  They disappear as the neutrino density
decreases and thus are only a transient phenomenon.

Our treatment leaves a number of interesting questions unresolved.
While the self-consistency conditions determine the static solution,
we have not realized under which circumstances such solutions
actually exist, although we have not found a counter example.

The adiabatic  solution is stable in typical numerical examples in
that the true solution of the EOMs tracks this adiabatic solution.
There could be situations where this is not the case, leading to
kinematical decoherence rather than a collective oscillation of the
individual modes.

Even though several important aspects of this system are only
numerically observed and not analytically proven, we believe that our
adiabatic treatment of the EOMs nicely explains many of the
intriguing features of this nonlinear system.


\begin{acknowledgments}
This work was partly supported by the Deutsche Forschungsgemeinschaft
under grant No.~TR-27, by the Cluster of Excellence ``Origin and
Structure of the Universe'' (Munich and Garching), and by the
European Union under the ILIAS project, contract
No.~RII3-CT-2004-506222. A.S.~acknowledges support by the Alexander
von Humboldt Foundation and hospitality by the Max-Planck-Institut
f\"ur Physik during a visit where this work was begun.
\end{acknowledgments}


\end{document}